%% file: narison193.tex
\documentclass[3p,times,twocolumn]{elsarticle}
 \biboptions{comma,sort&compress}
 
\usepackage{graphicx}
\usepackage{here}
\usepackage{ecrc}
\volume{00}

\firstpage{1}

\journalname{Nuclear and Particle Physics Proceedings}

\runauth{Stephan Narison}

\jid{nppp}

\jnltitlelogo{Nuclear and Particle Physics Proceedings}

\usepackage{amssymb}
\usepackage[figuresright]{rotating}

\def\beq{\begin{equation}}
\def\eeq{\end{equation}}
\def\bea{\begin{eqnarray}}
\def\eea{\end{eqnarray}}
\def\bq{\begin{quote}}
\def\eq{\end{quote}}

\def\nnb{\nonumber}
\def\ga{\left(}
\def\dr{\right)}

\def\lb{\lbrack}

\def\lrar{\Longrightarrow}
																
\def\nnb{\nonumber}
\def\la{\langle}
\def\ra{\rangle}
\def\nin{\noindent}
\def\ba{\vspace*{-0.2cm}\begin{array}}
\def\ea{\end{array}\vspace*{-0.2cm}}

\def\b{$\bullet~$}
\def\als{\alpha_s}

\def\gg2{ \la\alpha_s G^2 \ra}
\def\gg3{g^3f_{abc}\la G^aG^bG^c \ra}
\def\ggg4{\la\als^2G^4\ra}

\def\lb{\label}

\begin{document}

\begin{frontmatter}

%%
%%%%%%%%%%%%%%%%%%%%%%%%%%%%%%%%%%%%%%%%%%%%%%%%%
%\begin{document}
\title{
%$\la g^3f_{abc} G^aG^bG^c\ra$
%$\la g^3 f_{abc}G^3\ra$ 
% 
%$\alpha_s $, $\la \alpha_sG^2\ra$, $\overline{m}_{c,b}$ and $f_{B_c}$
%from  relativistic heavy quark sum rules$^*$} 
QCD parameters and $f_{B_c}$ from heavy quark sum rules$^*$} 
 
 \cortext[cor0]{Review talk presented at QCD19 (2-7 july 2019,
  Montpellier - FR) and at HEPMAD19 (14-20 october 2019, 
%  High-Energy Physics International Conferences, 
  Antananarivo-MG). }

 \author[label1]{Stephan Narison
 \corref{cor1} 
 }
   \address[label1]{Laboratoire
Univers et Particules , CNRS-IN2P3,  
Case 070, Place Eug\`ene
Bataillon, 34095 - Montpellier Cedex 05, France.}
\cortext[cor1]{ICTP-Trieste  high-energy physics consultant for Madagascar.}
\ead{snarison@yahoo.fr}

\pagestyle{myheadings}
\markright{ }
\begin{abstract}
\noindent
We report results of our recent works\,\cite{SN18,SN19} where we 
where the correlations  between the $c,b$-quark running masses $\overline{m}_{c,b}$, the gluon condensate $\la \alpha_s G^2\ra$ and the QCD coupling $\alpha_s$ in the $\overline{MS}$-scheme  from an analysis of the charmonium and bottomium spectra and the $B_c$-meson mass. We use optimized 
 ratios of relativistic Laplace sum rules (LSR) evaluated at the $\mu$-subtraction stability point  where higher orders PT and $D\leq 6-8$-dimensions non-perturbative condensates corrections 
 are included. 
 We obtain\,\cite{SN18} $\alpha_s(2.85)=0.262(9)$ 
 and $ \alpha_s(9.50)=0.180(8)$ from the (pseudo)scalar $M_{\chi_{0c(0b)}}-M_{\eta_{c(b)}}$ mass-splittings at $\mu=2.85(9.50)$ GeV. The most precise result from the charm channel leads to $\alpha_s(M_\tau)=0.318(15)$ and $\alpha_s(M_Z)=0.1183(19)(3)$ in excellent agreement with the world average: $\alpha_s(M_Z)=0.1181(11)$\,\cite{BETHKEa,PDG}.
 Updated results from a global fit of the (axial-)vector  and (pseudo)scalar channels using Laplace and Moments sum rules @ N2LO \,\cite{SN18} combined with the one from $M_{B_c}$\,\cite{SN19} lead to the {\it new tentative QCD spetral sum rules (QSSR) average :} $\overline{m}_c(\overline{m}_c)\vert_{\rm average}=  1266(6)$ MeV and  $\overline{m}_b(\overline{m}_b)\vert_{\rm average}=4196(8)$ MeV. 
 The values of the gluon condensate  $\la\alpha_s G^2\ra$ from the (axial)-vector charmonium channels combined with previous determinations in Table\,\ref{tab:g2}, leads to the {\it new QSSR average:}\,\cite{SN18}  $\la\alpha_s G^2\ra\vert_{\rm average}=(6.35\pm 0.35)\times 10^{-2}$ GeV$^4$. Our results clarify the (apparent) discrepancies between different estimates of $\la \alpha_s G^2\ra$ from $J/\psi$ sum rule but also shows the sensitivity of the sum rules on the choice of the $\mu$-subtraction scale. As a biproduct, we deduce the $B_c$-decay constants  $f_{B_c}=371(17)$ MeV and $f_{B_c}(2S)\leq 139(6)$ MeV. 
 
% which does not permit a high-precision estimate of $\overline{m}_{c,b}$ from the alone vector channel. 

%% keywords
%\ccode{Pac numbers: 11.55.Hx, 12.38.Lg, 13.20-Gd, 14.65.Dw, 14.65.Fy, 14.70.Dj}  
\end{abstract}
\end{frontmatter}
%\end{document}
%%%%%%%%%%%%%%%%%%%%%%%%%%%%%%%%%%
%\vspace*{-1.5cm}
\section{Introduction}
\vspace*{-0.25cm}
 \nin
%%%%%%%%%%%%%%%%%%%%%%%%%%%%%%%%%%%
Besides the importance of the QCD coupling $\overline{\alpha}_s$ and the running heavy quark masse $\overline{m}_{c,b}$, the non-perturbative
gluon condensates introduced by SVZ\,\cite{SVZa,SVZb,ZAKA} play important r\^ole in gluodynamics and in the
QCD spectral sum rules (QSSR) analysis where they enter as high-dimension operators in the OPE of the hadronic correlators. In particular, this is the case for the heavy quark systems and the 
pure Yang-Mills gluonia/glueball channels\,\cite{NSVZ,VENEZIA,SNG} where the light quark loops and condensates are absent to leading order. The heavy quark condensate contribution can be absorbed into the gluon one through the relation \cite{SVZa,SVZb}: 
\beq
\la \bar QQ\ra=-{\la\alpha_s G^2\ra/ (12\pi M_Q)}+....
\eeq
where a similar relation holds for the mixed heavy quark-gluon condensate $\la \bar QGQ\ra$ .  $G$ is the short hand notation for the gluon field strength $G^a_{\mu\nu}$ and $M_Q$ is the pole mass. 
The SVZ orignal value\,\cite{SVZa,SVZb}:
\beq
\la\alpha_s G^2\ra\simeq 0.04 ~{\rm GeV}^4~,
\label{eq:standard}
\eeq
extracted (for the first time) from charmonium sum rules \cite{SVZa,SVZb} has been challenged by different
authors (for reviews, see e.g \cite{SNB1,SNB2,SNB3,SNB4} and Table\,\ref{tab:g2}).  
One can see in Table\, \ref{tab:g2} that the results from standard SVZ and FESR sum rules for heavy and light quark systems vary in a large range but all of them are positive numbers, while the ones from analysis of the modified $\tau$-decays  moments allow negative values. 
However, one should notice from the original QCD expression of the $\tau$-decay rate\,\cite{BNPa,BNPb}  that the $\la\alpha_s G^2\ra$ gluon condensate contribution is absent to leading order indicating that the original $\tau$-decay rate is a bad place for extracting a such quantity\,\cite{SNTAU}.  
The presence of $\la\alpha_s G^2\ra$ in the analysis of \cite{DUFLOT,OPAL,ALEPH,DAVIER} is only an aritfact of the high-moments where the systematic errors needs to be better controlled. 
Earlier lattice calculations indicate a non-zero positive value of $\la\alpha_s G^2\ra$\,\cite{GIACOa,GIACOb,GIACOc,GIACOd}
while recent  estimates in Table\,\ref{tab:g2} give positive values but about  2-7 times higher than the phenomenological estimates. However, the subtraction of the perturbative contribution in the lattice analysis which is scheme dependent is not yet well-understood\,\cite{LEE}  such that a direct comparison of the lattice  results obtained at large orders of PT series with the ones from the truncated PT series used in the phenomenological analysis is quite delicate. These previous results indicate that  $\la\alpha_s G^2\ra$ is not yet well determined and motivate a reconsideration of its estimate. 

%The value of the dimension-four gluon condensate $\la\alpha_s G^2\ra$ havebeen extracted phenomenologically from different channels with the range of values given in Table\,\ref{tab:g2}.

A first step for the improvement of the estimate of the gluon condensate was the recent direct determination of 
the ratio of the dimension-six gluon condensate $\la g^3f_{abc} G^3\ra$ over the dimension-four one $\la\alpha_s G^2\ra$ from the heavy quark systems with the value\,\cite{SNcb1,SNcb2,SNcb3}:
\beq
\rho\equiv \la g^3f_{abc} G^3\ra/ \la \alpha_s G^2\ra=(8.2\pm 1.0)~{\rm GeV}^2,
\label{eq:rcond}
\eeq
which differs significantly from the instanton liquid model estimate\,\cite{NIKOL2,SHURYAK,IOFFE2} and may question the validity of the instanton liquid model approximation. 
Earlier lattice results in pureYang-Mills found:  $\rho\approx 1.2$ GeV$^2$\,\cite{GIACOa,GIACOb,GIACOc,GIACOd} such that it is important to have new lattice results for  this quantity. Note however, that the value given in Eq.\,\ref{eq:rcond} might also be an effective
value of the unknown high-dimension condensates not taken into account in the analysis of \,\cite{SNcb1,SNcb2,SNcb3} when requiring the fit of the data by the truncated OPE at that order in the extreme case where the OPE does not converge. We shall see that the effect of the $ \la g^3f_{abc} G^3\ra$ term is a small correction at the stability  region where the optimal results are extracted. 

 In this paper, we pursue a such program by reconsidering the extraction of the  lowest dimension QCD parameters from the (axial-)vector and (pseudo)scalar charmonium and bottomium spectra taking into account the correlations between $\alpha_s$, the gluon condensate $\la \alpha_s G^2\ra$,  and the $c,b$-quark running masses. We shall use these parameters for predicting the known masses of the (pseudo)scalar heavy quarkonia ground states and also re-extract $\alpha_s$ and $\la \alpha_s G^2\ra$ from the mass-splittings $M_{\chi_{0c(0b)}}-M_{\eta_{c(b)}}$.  In so doing, we shall work with the example of the QCD Laplace sum rules (LSR) where the corresponding Operator Product Expansion (OPE) in terms of condensates is more convergent than the moments evaluated at small momentum. 
% \vspace*{-0.2cm}
 %%%%%%%%%%%%%%%%%%%%%
\section{The heavy quarkonia Laplace sum rules (LSR)}
%%%%%%%%%%%%%%%%%%%%%%
\vspace*{-1.5cm}
 %%%%%%%%%%%%%%%%%%%%%%%%%%%%%%%%%%%%%%%
 {\scriptsize
\begin{table}[H]
\begin{center}
%\begin{table*}[hbt]
\setlength{\tabcolsep}{0.15pc}
%\newlength{\digitwidth} \settowidth{\digitwidth}{\rm 0}
%\catcode`?=\active \def?{\kern\digitwidth}

 \caption{\footnotesize    
Selected determinations of  $\la \alpha_s G^2\ra$ in units of  $10^{2}$ [GeV$^4$] 
from charmonium, bottomium and light quark sum rules (SR).  The numbers marked with * are not included in the average. This average take into account the new results obtained in\,\cite{SN18} from Exponential / Laplace Sum Rules (LSR). Estimates from variants of the SVZ  sum rules using some weight functions are not considered here. The ones from high-moments of  $\tau-$decays and from the lattices are only mentioned for comparisons. }
%\vspace*{-0.5cm}
    {\footnotesize
\begin{tabular}{llllccc}
%\hline
\hline
%\\
Sources&$\la \alpha_s G^2\ra$&References \\
%$\la \alpha_s G^2\ra$ &$\sqrt{R^{th}_{J/\psi}}$&$\sqrt{R^{exp}_{J/\psi}}$&$M^{th}_{\eta_c}$&$M^{exp}_{\eta_c}$\\
%\\
\hline
\\
\multicolumn{3}{l}{\bf\boldmath {\bf Vector Charmonium SR}} & \\
$q^2=0$-moments & $4\pm 2$ &SVZ 79\,\cite{SVZa,SVZb} (guessed error)\\
$q^2\not=0$-moments&$5.3\pm 1.2$&RRY 81-85\,\cite{RRY}\\
--&$9.2\pm 3.4$& Miller-Olssson 82\,\cite{OLSSON}\\
--&$\approx 6.6^*$&Broadhurst et al. 94\,\cite{BAIKOV}\\
-- &$2.8\pm 2.2 $& Ioffe-Zyablyuk 07\,\cite{IOFFEa,IOFFEb}\\
--& $7.0\pm 1.3$& Narison 12a\,\cite{SNcb2}\\
LSR &$12\pm 2$&Bell-Bertlmann 82\,\cite{BELLa,BELLb,BERTa,BERTb,BERTc,BERTd,NEUF} \\
--&$17.5\pm 4.5$&Marrow et al. 87\,\cite{SHAW}\\
--& $7.5\pm 2.0$& Narison 12b\,\cite{SNcb3}\\
%-- & $7.4\pm 2.2$& {\it (Axial)-Vector: This work }\\

\\
\multicolumn{2}{l}{\bf\boldmath {\bf Vector Bottomium SR}} & \\
Non-rel. vector mom.&$5.5\pm 3.0$&Yndurain 99\,\cite{YND}\\
\\
\multicolumn{3}{l}{\bf\boldmath {\bf Other Charmonium and Bottomium SR}} & \\
LSR  $M_\psi-M_{\eta_c}$&$10\pm 4$&Narison 96\,\cite{SNHeavy,SNHeavy2}\\
--\hspace*{0.35cm} $M_{\chi_b}-M_\Upsilon$ &$6.5\pm 2.5$&Narison 96\, \cite{SNHeavy,SNHeavy2}\\
--\hspace*{0.35cm} $M_\psi\oplus M_{\chi_{c1}}$&$8.5\pm 3.0$&Narison 18\,\cite{SN18}\\
--\hspace*{0.35cm} $ M_{\chi_{0c,0b}}\hspace*{-0.1cm}-M_{\eta_{c,b}}$&$6.39\pm 0.35$&Narison 18\,\cite{SN18}\\
\\
\multicolumn{2}{l}{\bf\boldmath {\bf $e^+e^-\to$ I=1 Hadrons SR}} & \\
LSR & $0.9\sim 6.6^*$  &Eidelman et al. 79\,\cite{EID}\\
Ratio of LSR &$4\pm1$ &Launer et al. 84\,\cite{LNT}  \\
FESR &$13\pm 6$ & Bertlmann et al. 88\,\cite{PEROTTETa,PEROTTETb}\\
Infinite norm &$1 \sim 30^*$ & Causse-Mennessier\,\cite{MENES}\\
$\tau$-like decay & $7\pm 1$ & Narison 95\,\cite{SNIa,SNIb}\\
\\
\multicolumn{3}{l}{\bf\boldmath {\bf $\tau-$decay SR}} & \\
Axial spectral function&$6.9\pm 2.6$& Dominguez-Sola 88\,\cite{SOLA}\\
\\
%\multicolumn{2}{l}
%\hline
%\multicolumn{2}{l}
{\bf\boldmath {\bf SR Average 2018}} &\bf\boldmath{$6.35\pm 0.35$} &\\
%\\
\hline
\\
\multicolumn{3}{l}{\bf\boldmath {\bf $\tau-$decay with high moments SR}} & \\
ALEPH collaboration &$6.3\pm 1.2$ & Duflot 95\,\cite{DUFLOT}   \\
CLEO II collaboration& $2.4\pm1.0$& Duflot 95\,\cite{DUFLOT} \\
OPAL collaboration &$-0.9\sim +4$& Ackerstaff et al. 99\,\cite{OPAL}\\
ALEPH collaboration& $-5\sim +6$& Schael et al. 05\,\cite{ALEPH} \\
ALEPH  collaboration&$-12\sim -0.6$& Davier et al. 14\,\cite{DAVIER} \\
%\\
\hline
\\
\multicolumn{3}{l}{\bf\boldmath {\bf Lattice}} & \\
{\cal O}($\alpha_s^{12}$) & $\approx 13$& Rakow 05\,\cite{RAKOW,BURGIO,HORLEY}\\
{\cal O}($\alpha_s^{35}$) & $\approx 27$& Bali-Pineda 15\,\cite{BALIa,BALIb}\\
Average plaquette &$\approx 44$&Lee 14\,\cite{LEE}\\

%\\
\hline
%\hline
\end{tabular}
%\end{tabular*}
}
\label{tab:g2}
\end{center}
\vspace*{-0.5cm}
\end{table}
}
%%%%%%%%%%%%%%%%%%%%%%%%%%%%%
\subsection*{\b Form of the sum rule}
%%%%%%%%%%%%%%%%%%%%%%%%%%%%% 
%%%%%%%%%%%%%%%%%%%%%%%%%%%%%%%%%%%%%%
We shall work with the  Finite Energy version of the QCD Laplace sum rules (LSR) and their ratios:
 \beq
\hspace*{-0.5cm} {\cal L}^c_n\ga \tau\dr=\int_{4m_Q^2}^{t_c}\hspace*{-0.2cm} dt ~t^n~e^{-t\tau}{\rm Im}\Pi_{V(A)}(t)~,~
 {\cal R}^c_n(\tau)=\frac{{\cal L}^c_{n+1}} {{\cal L}^c_n}~,
 \lb{eq:mom}
 \eeq
 where $\tau$ is the LSR variable, $t_c$ is  the threshold of the ``QCD continuum" which parametrizes, from the discontinuity of the Feynman diagrams, the spectral function  ${\rm Im}\Pi_{V(A)}(t,m_Q^2,\mu)$   associated to the transverse part  $\Pi_{V(A)}(q^2,m_Q^2,\mu)$ of the two-point correlator: 
  \bea
&&\hspace*{-1.2cm}\Pi^{\mu\nu}_{V(A)}(q^2) \equiv  i \hspace*{-0.15cm}\int \hspace*{-0.15cm}d^4x ~e^{\rm- iqx}\la 0\vert {\cal T} J^\mu_{V(A)}(x)\ga J^\nu_{V(A)}(0)\dr^\dagger \vert 0\ra\nnb\\
&&\hspace*{-0.5cm}
  = -\ga g^{\mu\nu}q^2-q^\mu q^\nu\dr \Pi_{V(A)}(q^2)+q^\mu q^\nu \Pi^{(0)}_{V(A)} (q^2),
\label{eq:2-point}
 \eea
 where : $J_{V(A)}^\mu(x)=\bar Q \gamma^\mu(\gamma_5) Q(x)$ is the heavy quark local vector (axial-vector) current.  In the  (pseudo)scalar channel associated  to the local current $J_{S(P)}=\bar Q i(\gamma_5) Q(x)$, we work with the correlator:
 \beq
\hspace*{-0.3cm} \Psi_{S(P)}(q^2)=i \hspace*{-0.15cm}\int {\hspace*{-0.15cm}d^4x} ~e^{-iqx}\la 0\vert {\cal T} J_{S(P)}(x)\ga J_{S(P)}(0)\dr^\dagger \vert 0\ra,
 \label{eq:2-pseudo}
 \eeq
 which is related to the longitudinal part $\Pi^{(0)}_{V(A)}(q^2)$ of the (axial-)vector one through  the Ward identity\,\cite{SNB1,SNB2,BECCHI}:
 \beq
q^2 \Pi^{(0)}_{A(V)}(q^2)=\Psi_{P(S)}(q^2)-\Psi_{P(S)}(0)~.
 \eeq
Working with $\Psi_{P(S)}(q^2)$ is safe as $\Psi_{P(S)}(0)$ should affect the $Q^2$-moments and the exponential sum rules derived from $\Pi^{(0)}_{A(V)}(q^2)$ which is not accounted for in e.g \,\cite{RRY,BERTb,SHAW} . 
 
 Originally named Borel sum rules by SVZ because of the appearance of  a factorial suppression factor in the non-perturbative condensate contributions into the OPE, it has been shown by \cite{SNR} that the PT radiative corrections satisfy instead the properties of an inverse Laplace sum rule though the present given name here. 

%  from which we shall extract independently some of the previous QCD parameters. 
%%%%%%%%%%%%%%%%%%%%%%%%%%%%%
\subsection*{\b Parametrisation of the spectral function}
%%%%%%%%%%%%%%%%%%%%%%%%%%%%% 
${\rm Im}\Pi_V(t)$ is  related to  the ratio  $R_{e^+e^-}$ of the total cross-section of $\sigma(e^+e^-\to$ hadrons) over $\sigma(e^+e^-\to \mu^+\mu^-)$ through the optical theorem. Expressed in terms of the leptonic widths and meson masses, it reads in a 
narrow width approximation (NWA):
 \bea
R_{e^+e^-}&\equiv& 12\pi{\rm Im} \Pi_V(t)\nnb\\
 &=&\frac{9\pi}{ Q_V^2 \alpha^2}\sum_{}M_{V}\Gamma_{V\to e^+e^-}\delta(\ga t-M_V^2\dr,
 \eea
 where $M_{V}$ and $\Gamma_{V\to e^+e^-}$ are the mass and leptonic width of the $J/\psi$ or $\Upsilon$ mesons; $Q_V=2/3 (-1/3)$ is the charm (bottom) electric charge in units of $e$; $\alpha=1/133$ is the running electromagnetic coupling evaluated at $M^2_{V}$. We shall use the experimental values of the $J/\psi$ and $\Upsilon$ parameters compiled by PDG\,\cite{PDG}. We include the contributions of the $\psi(3097)$ to $\psi(4415)$ and $\Upsilon(9460)$ to $\Upsilon(11020)$ within NWA. 
The high-energy part of the spectral function is parametrized by the ``QCD continuum" from a threshold $t_c$ (we use $\sqrt{t_c^c}=4.6$ GeV and $\sqrt{t_c^b}$= 11.098 GeV just above the last resonance).

 In the case of the axial-vector and (pseudo)scalar channels where there are no complete data, we use the duality ansatz:
 \bea
\hspace*{-0.5cm} {\rm Im} \{\Pi(t);\,\Psi(t)\}&\simeq& f_H^2 M_H^{\{0;2\}} \delta(t-M_H^2) +\nnb\\
 &&\Theta(t-t_c) ``QCD~{\rm  continuum}",
 \eea
 where $M_H$ and $f_H$ are the lowest ground state mass and coupling analogue to $f_\rho$ and $f_\pi$. This implies : 
 \beq
  {\cal R}^{c}_n\equiv {\cal R}\simeq M_H^2~,
  \eeq
 indicating that the ratio of moments appears to be a useful tool for extracting the masses of hadrons \cite{SNB1,SNB2,SNB3,SNB4}. We shall work with the lowest ratio of moments $ {\cal R}^{c}_0$.
Exponential sum rules have been used successfully by SVZ for light quark systems \cite{SVZa,SVZb,SNB1,SNB2,SNB3,SNB4} and extensively by Bell and Bertlmann for heavy quarkonia in their relativistic and non-relativistic versions \cite{BELLa,BELLb,BERTa,BERTb,BERTc,BERTd,NEUF,SHAW,SNHeavy,SNHeavy2}.  
%%%%%%%%%%%%%%%%%%%%%%%%%%%%%
\subsection*{\b QCD Perturbative expressions @N2LO}
%%%%%%%%%%%%%%%%%%%%%%%%%%%%%
The perturbative QCD expression of the vector channel is deduced from the well-known spectral function to order $\alpha_s$ within  the on-shell renormalization scheme \cite{KALLEN,SCHWINGER}.  The one of the axial-vector current has been obtained in\,\cite{SCHILCHER,BROAD,GENER,RRY}. To order $\alpha_s^2$ (N2LO), the spectral functions are usually parametrized as:
\bea
R^{(2)}&\equiv &C_F^2R^{(2)}_A+C_AC_FR^{(2)}_{NA}+C_FT_Qn_lR^{(2)}_l\nnb\\
&&+C_FT_Q(R^{(2)}_F+ R^{(2)}_S+R^{(2)}_G)~,
\eea
which are respectively the abelian (A), non-abelian (NA), massless (l) and heavy (F) internal quark loops,  singlet (S) and double bubble gluon (G) contributions. $C_F=4/3,$ $ C_A=3, T_Q=1/2$ are usual SU(3) group factors and $n_l$ is the number of light quarks. 
We use the (approximate) but complete result in the on-shell scheme given by\,\cite{CHET0} for the abelian and non-abelian contributions.  The one from light quarks comes from\,\cite{TEUBNER,HOANGa,CHET1}. The one from heavy fermion internal loop comes from\,\cite{CHET2} for the vector current while the one from the axial current is (to our knowledge) not available.
 The singlet one due to double triangle loop comes from\,\cite{CHET3}. The one from the gluonic double-bubble reconstructed from massless fermions comes from\,\cite{TEUBNER,HOANGa,CHET2}.   
 The previous on-shell expressions are transformed into the $\overline{MS}$-scheme through the relation between the on-shell $M_Q$ and running $ \overline{m}_Q(\mu)$ quark masses\,\cite{SNB1,SNB2,SNB3,SNB4,TAR,COQUEa,COQUEb,SNPOLEa,SNPOLEb,BROAD2a,AVDEEV,BROAD2b,CHET2a,CHET2b} @N2LO:
\bea
\hspace*{-0.15cm}M_Q &=& \overline{m}_Q(\mu)\Big{[}
1+\frac{4}{3} a_s+ (16.2163 -1.0414 n_l)a_s^2\nnb\\
&&+{\rm ln} \ga a_s+(8.8472 -0.3611 n_l) a_s^2\dr
\nnb\\ &&
+{\rm ln}^2\ga 1.7917 -0.0833 n_l\dr a_s^2+\cdots\Big{]},
\label{eq:pole}
\eea
for $n_l$ light flavours where $\mu$ is the arbitrary subtraction point and $a_s\equiv \alpha_s/\pi$, ln$\equiv \ln{\ga{\mu}/{ M_Q}\dr^2}$.
%%%%%%%%%%%%%%%%%%%%%%%%%%%%%
\subsection*{\b QCD Non-Perturbative expressions @LO}
%%%%%%%%%%%%%%%%%%%%%%%%%%%%%
Using the OPE \`a la SVZ, the non-perturbative contributions to the two-point correlator can be parametrized by the sum of higher dimension condensates:
\beq
{\rm Im} \Pi(t)=\sum C_{2n}(t,m^2,\mu^2))\la O_{2n}\ra~: n=1,2,...
\eeq
where $C_{2n}$ are Wilson coefficients calculable perturbatively and $ \la O_{2n}\ra$
are non-perturbative condensates.  In the exponential sum rules, the order parameter  is the sum rule variable $\tau$ while for the heavy quark systems
the relevant condensate contributions at leading order in $\alpha_s$ are the gluon condensate $\la \alpha_s G^2\ra$ of dimension-four\,\cite{SVZa,SVZb},  the dimension-six gluon $ \la g^3f_{abc} G^3\ra $ and light four-quark  $\alpha_s\la\bar uu\ra^2 $ condensates\, \cite{NIKOLa,NIKOLb}. The condensates of dimension-8 entering in the sum rules are of seven types\,\cite{NIKOL2}. They can be expressed in different basis depending on how each condensate is estimated (vacuum saturation\,\cite{NIKOL2} or modified vacuum  saturation\,\cite{BAGAN}).  Our estimate of these D=8 condensates is the same as in\,\cite{SNcb2}. For the vector channel, we use the analytic expressions of the different condensate contributions  given by Bertlmann\,\cite{BERTb}. We shall not include the eventual $D=2$ coperator induced by a tachyonic gluon mass\,\cite{ZAK1,CNZ1} as
it is dual to the contribution of large order terms\,\cite{SNZ}, which we estimate using a geometric growth of the PT series. In various examples, its contribution is numerically negligible\,\cite{SND2}.
%%%%%%%%%%%%%%%%%%%%%%%%%%%%%%%%%
\subsection*{\b Initial QCD input parameters }
%%%%%%%%%%%%%%%%%%%%%%%%%%%%%%%%%
In the first iteration, we shall use the following QCD input parameters (mass in units of MeV):
\bea
&&\hspace*{-1.25cm}\alpha_s(M_\tau)=0.325^{+0.008}_{-0.016}~ ,~\la\alpha_s G^2\ra\,\,= (0.07\pm 0.04) ~{\rm GeV}^4.\nnb\\
&&\hspace*{-1.25cm}\overline{m}_c(\overline{m}_c)= (1261\pm 17)~,~
%~{\rm MeV},
\overline{m}_b(\overline{m}_b)= (4177\pm 11)
%~{\rm MeV}
~,
\label{eq:param}
\label{eq:mcmom}
\eea
The central value of $\alpha_s$ comes from $\tau$-decay\,\cite{SNTAU,PICHTAUb}. The range covers the one allowed by PDG\,\cite{PDG,BETHKEa} (lowest value) and the one from our determination from $\tau$-decay (highest value)\,\cite{SNTAU}. The values of $\overline{m}_{c,b}(\overline{m}_{c,b})$ are the average  from our recent determinations from charmonium and bottomium sum rules \,\cite{SNcb1,SNcb2}.
%:
%\beq
%\overline{m}_c(\overline{m}_c)=  (1261\pm 17)~{\rm MeV}~,
%\eeq
% which we have enlarged by taking into account the range allowed by PDG\,\cite{PDG}.
%one of $\overline{m}_{c,b}$ from our recent analysis of the $J/\psi$ and $\Upsilon$ sum rules\,\cite{SNcb1, SNcb2, SNcb3}. In these cases, the error bar takes into account the range allowed  by PDG\,\cite{PDG,BETHKEa,BETHKEb,BETHKEc}. 
The value of  $\la\alpha_s G^2\ra$  almost covers the range from different determinations mentioned in Table\,\ref{tab:g2} and reviewed in\,\cite{SNB1,SNB2,SNHeavy,SNHeavy2}. We shall use the ratio of condensates given in Eq.\,\ref{eq:rcond}. For the light four-quark condensate, we shall use the value:
\beq
 \alpha_s\la\bar uu\ra^2=(5.8\pm 1.8)\times 10^{-4}~{\rm GeV}^6~,
 \eeq
  obtained from the original $\tau$-decay rate\,\cite{SNTAU}  where the gluon condensate does not contribute to LO\,\cite{BNPa,BNPb} and by some other authors from the light quark systems\,\cite{SNB1,SNB2, LNT,JAMI2a,JAMI2b,JAMI2c} where a violation by a factor about 3--4 of the vacuum saturation assumption has been found. 
%%%%%%%%%%%%%%%%%%%%%%%%%%%%%%%%%
\section{Charmonium Ratio of  LSR Moments $ {\cal R}_{J/\psi(\chi_{c1})}$  }
%%%%%%%%%%%%%%%%%%%%%%%%%%%%%%%%%
%%%%%%%%%%%%%%%%%%%%%%%%%%%%%%%%%
\subsection*{\b Convergence of the  PT series}
%%%%%%%%%%%%%%%%%%%%%%%%%%%%%%%%%
%%%%%%%%%%%%%%%%%%%%%%%%%%%%%%%%%%%%%%%
\begin{figure}[hbt]
\begin{center}
\hspace*{-6cm}{\bf a)}\\
\includegraphics[width=10.5cm]{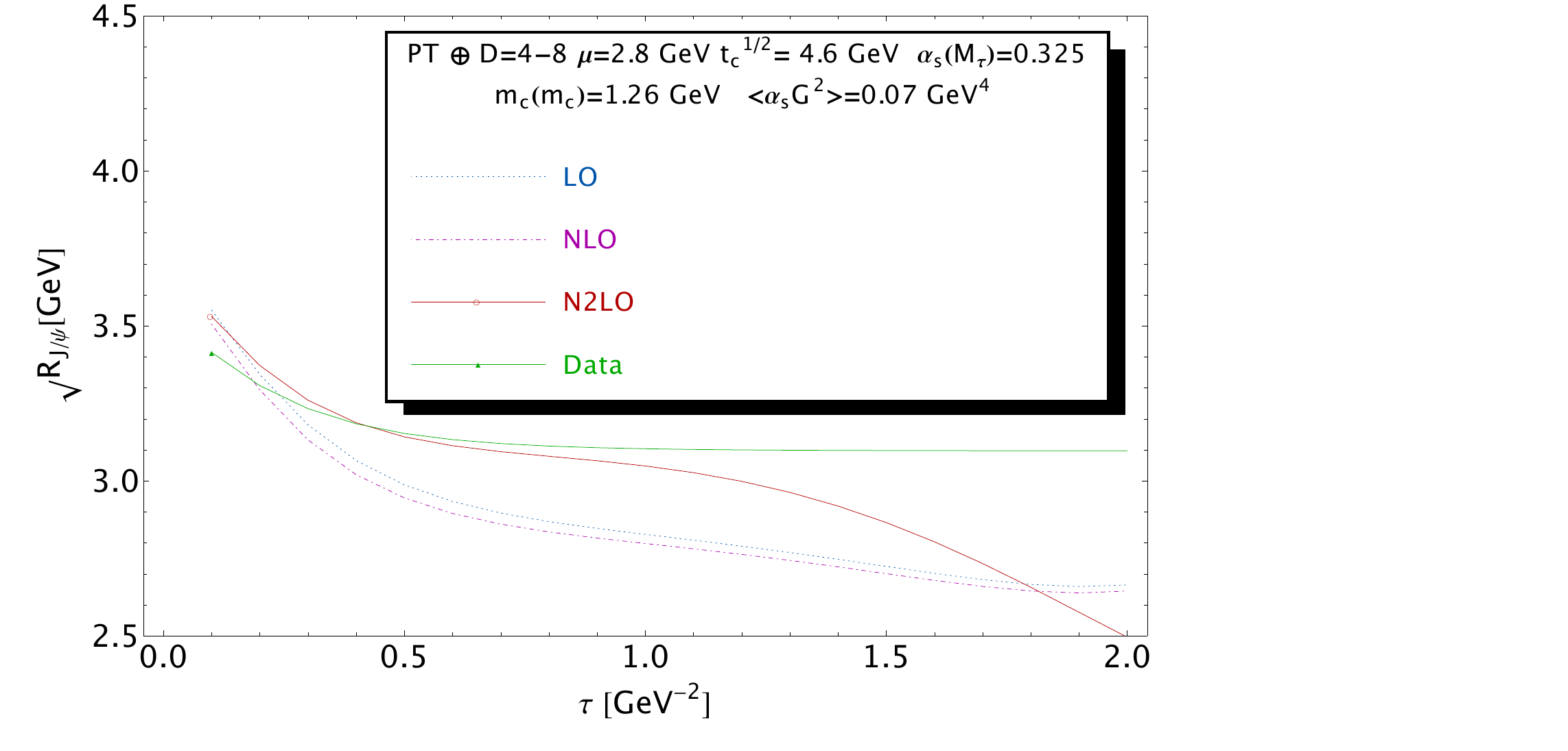}\\
\hspace*{-6cm}{\bf b)}\\
\includegraphics[width=10.5cm]{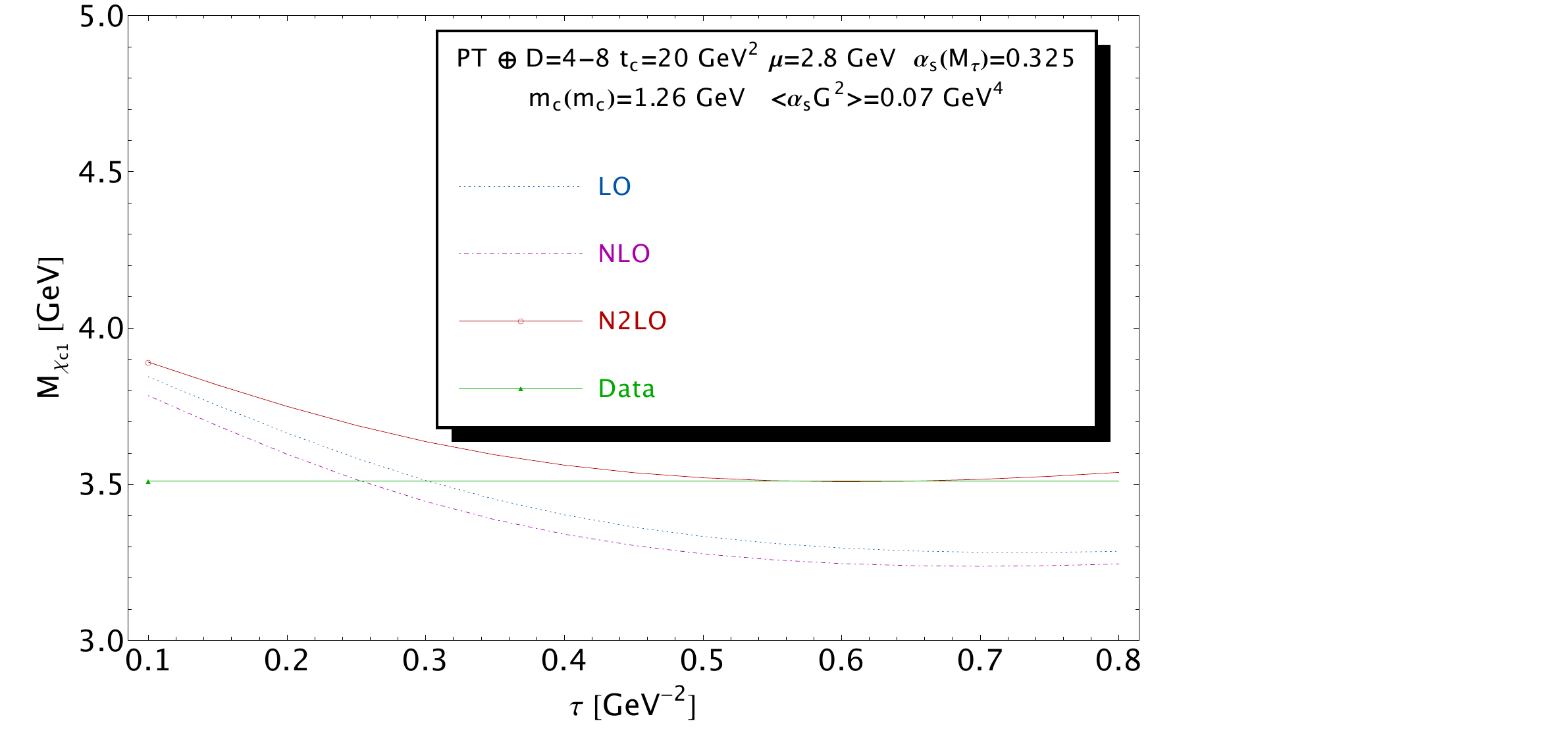}
\vspace*{-0.5cm}
\caption{\footnotesize  Behaviour of the ratio of moments ${\cal R}$ versus $\tau$ in GeV$^{-2}$
at different orders of perturbation theory. The input and the meaning of each curve are given in the legends:  {\bf a)}  $J/\psi$ and {\bf b)}  $\chi_{c1}$.} 
\label{fig:pertc}
\end{center}
\vspace*{-0.5cm}
\end{figure} 
%%%%%%%%%%%%%%%%%%%%%%%%%%%%%%%%%%%%%%%%%
In so doing, we shall work with the renormalized (but non-resummed renormalization group) perturbative (PT) expression where the subtraction point $\mu$ appears explicitly. We include  the known N2LO terms. The $D=8(6)$ condensates contributions are included for the (axial-)vector current. The value of $\sqrt{t_c}=4.6$ GeV is chosen just above the $\psi(4040)$ mass for the vector current where the sum of all lower mass $\psi$ state contributions are included in the spectral function. For the axial current, we use (as mentioned) the duality ansatz and leave $t_c$ as a free parameter which we shall fix after an optimisation of the sum rule.  We evaluate the ratio of moments at $\mu=2.8$ GeV and for a given value of $t_c=20$ GeV$^2$ for the $\chi_{c1}$ around which they will stabilize (as we shall show later on). The analysis is illustrated in Fig.\,\ref{fig:pertc}.
On can notice the importance of  the N2LO contribution which is dominated by the abelian and non-abelian contributions.
%\,\footnote{Approximate expressions expanded in terms of $m^2/t$ at order $\alpha_s^3$ are not included here but seem to be quite large and require a more complete evaluation. }. 
The N2LO effects go towards the good direction of the values of the experimental masses. 
%%%%%%%%%%%%%%%%%%%%%%%%%%%%%%%%%
\subsection*{\b LSR variable $\tau$-stability and Convergence of the  OPE }
\begin{figure}[hbt]
\begin{center}
\hspace*{-6cm}{\bf a)}\\
\includegraphics[width=10cm]{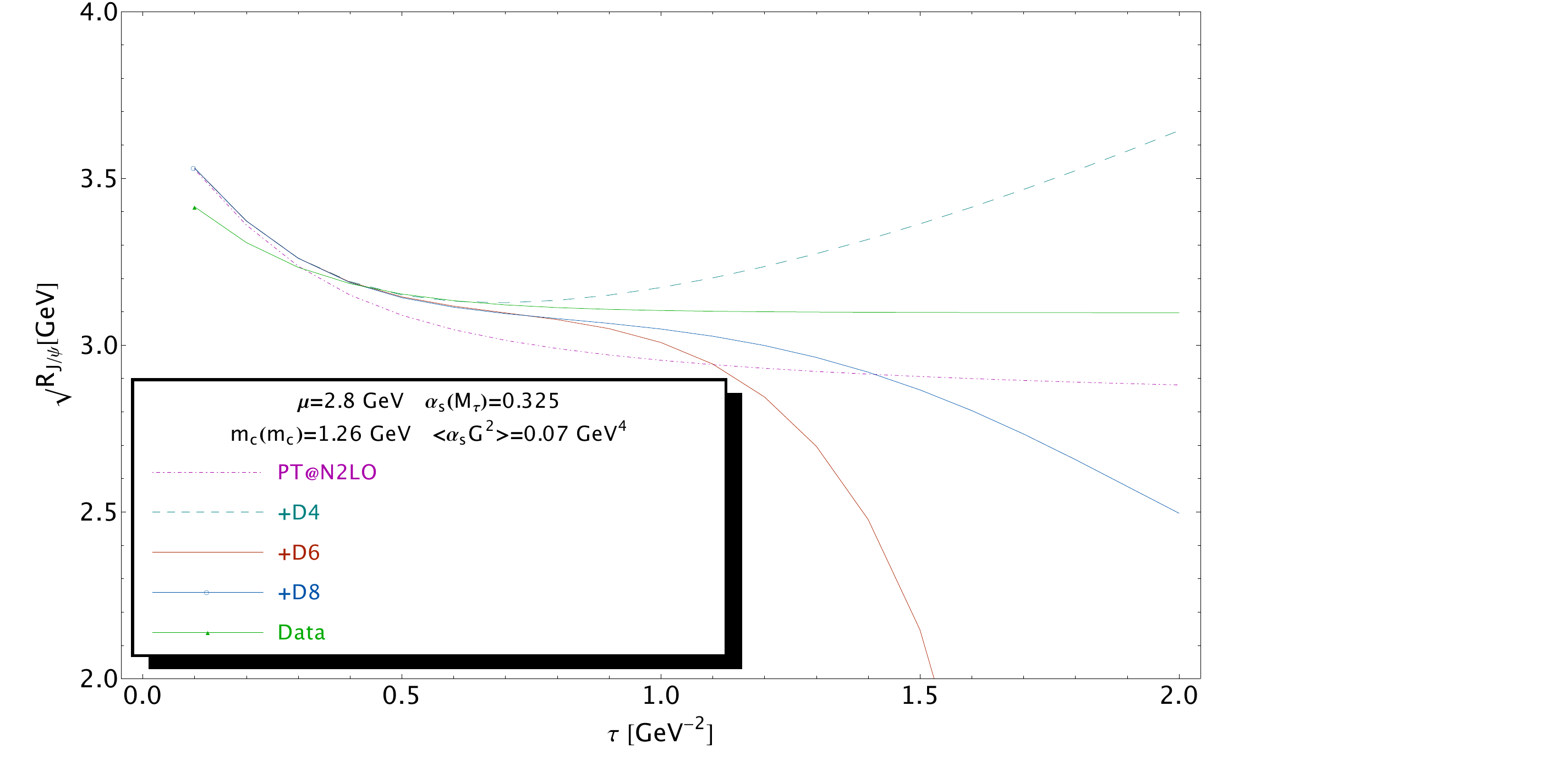}\\
\hspace*{-6cm}{\bf b)}\\
\includegraphics[width=10.5cm]{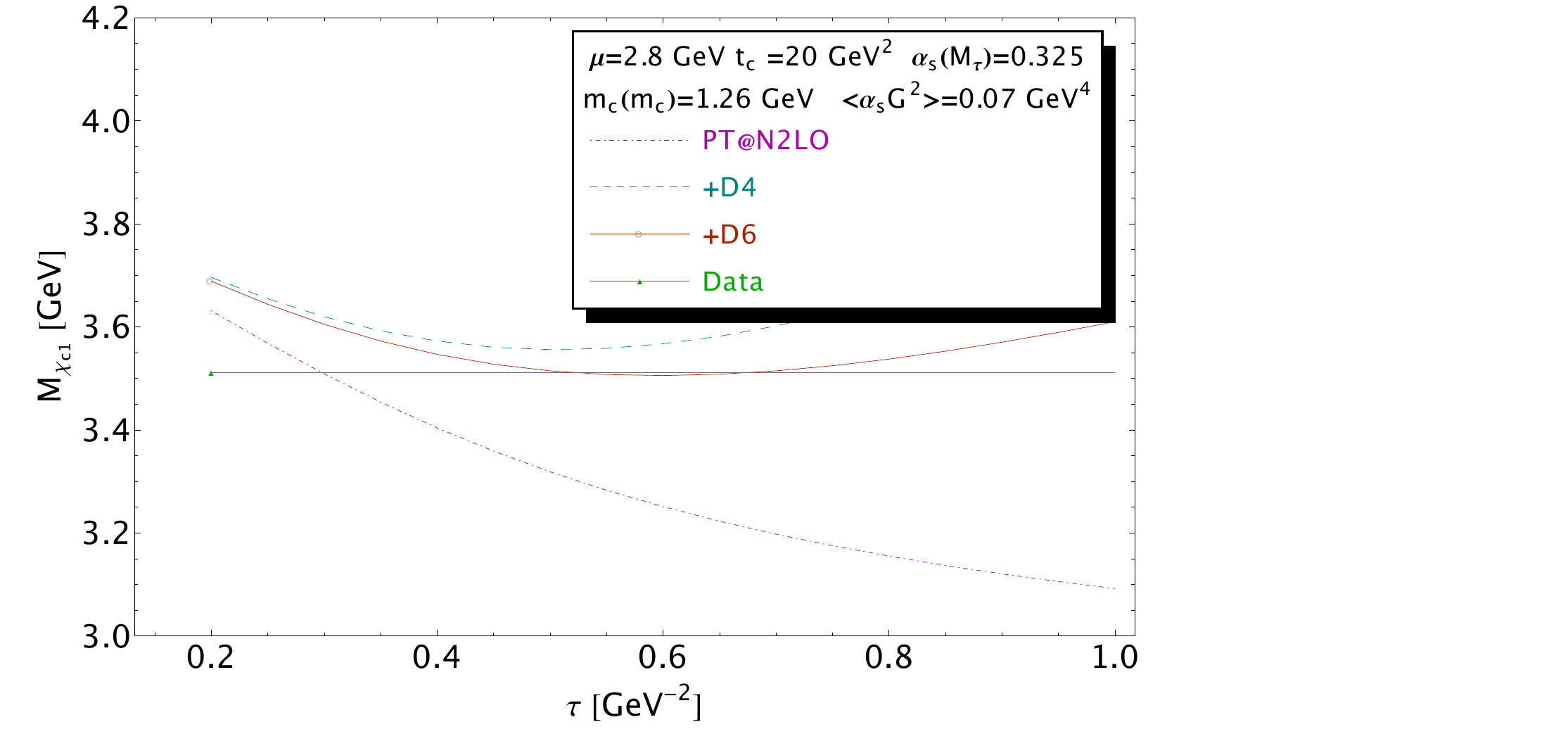}
\vspace*{-0.5cm}
\caption{\footnotesize  The same as in Fig.\,\ref{fig:pertc} but 
for  different truncation of the OPE:  {\bf a)}  $J/\psi$ and {\bf b)}  $\chi_{c1}$.} 
\label{fig:npertc}
\end{center}
\vspace*{-0.5cm}
\end{figure}

%%%%%%%%%%%%%%%%%%%%%%%%%%%%%%%%%
The OPE is done in terms of the exponential sum rule variable $\tau$. We show in Fig.\,\ref{fig:npertc} the effects of the condensates of different dimensions. One ca notice that the presence of condensates are  vital for having $\tau$-stabilities which are not there for the PT-terms alone. The $\tau$-stability is reached for $\tau\simeq 0.6$ GeV$^{-2}$. At a given order of the PT series, the contributions of the $D=8$ condensates are negligible at the $\tau$-stability region while the $D=6$ contribution goes again to the right track compared with  the data. 
%%%%%%%%%%%%%%%%%%%%%%%%%%%%%%%%%%%%%%%
%%%%%%%%%%%%%%%%%%%%%%%%%%%%%%%%%%%%%%%%%
%%%%%%%%%%%%%%%%%%%%%%%%%%%%%%%%%%%%%%%%%
\begin{figure}[hbt]
\begin{center}
\includegraphics[width=10cm]{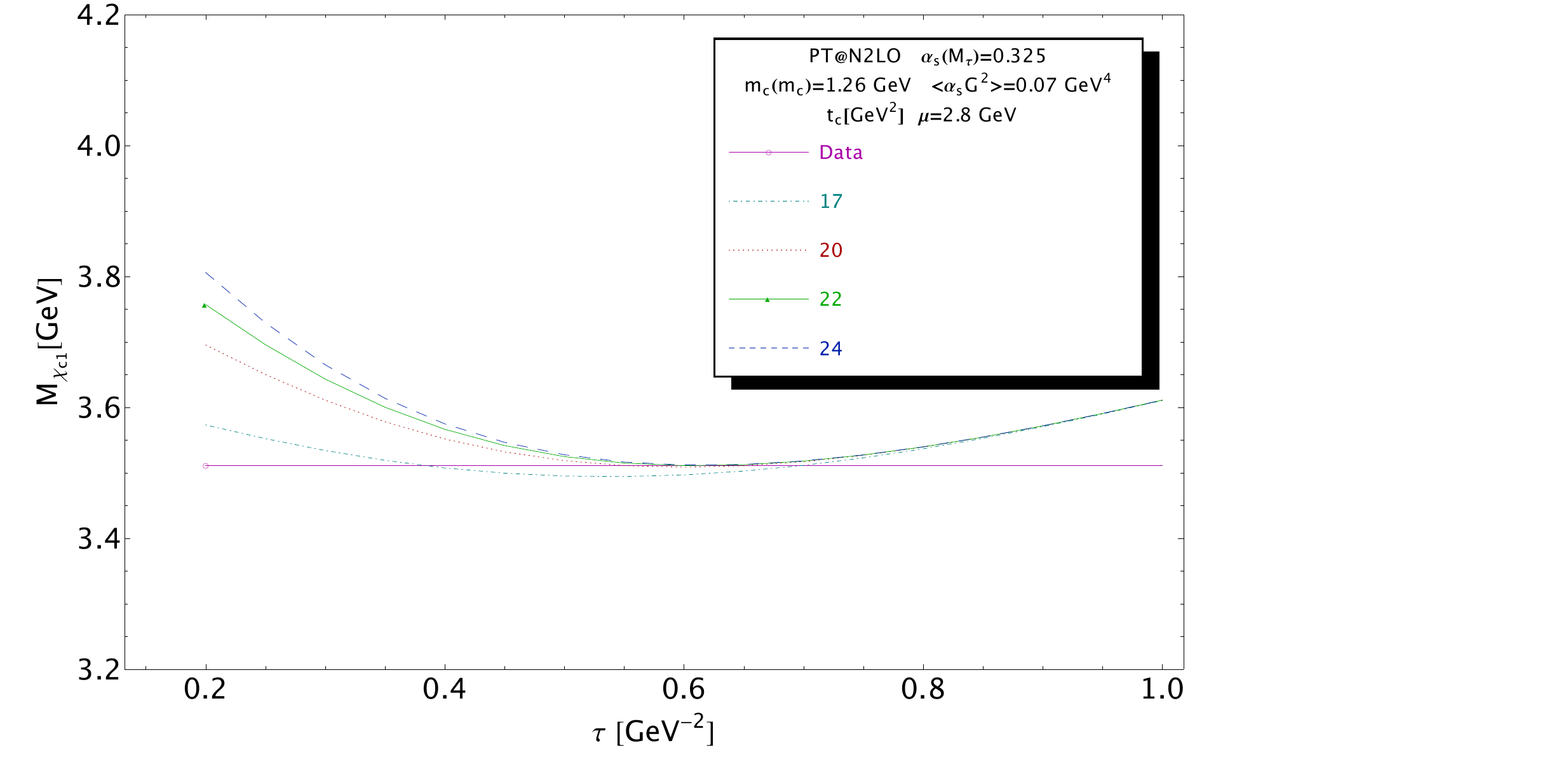}
\vspace*{-0.5cm}
\caption{\footnotesize  Behaviour of the ratio of moments ${\cal R}_{\chi_{c1}}$ versus $\tau$ in GeV$^{-2}$. The input and the meaning of each curve are given in the legend.} 
\label{fig:chi1c-tau}
\end{center}
\vspace*{-0.5cm}
\end{figure} 
%%%%%%%%%%%%%%%%%%%%%%%%%%%%%%%%%%%%%%%%%
%%%%%%%%%%%%%%%%%%%%%%%%%%%%%%%%%%%%%%%%%
\subsection*{\b Continuum threshold $t_c$-stability for  $ {\cal R}_{\chi_{c1}}$  }
%%%%%%%%%%%%%%%%%%%%%%%%%%%%%%%%%%%%%%%%%
We show the analysis in Fig.\,\ref{fig:chi1c-tau} where the curves correspond to different $t_c$-values. We find nice $t_c$-stabilities where we take the value :
\beq
t_c\simeq (17\sim 22) ~{\rm GeV}^2~,
\label{eq:tc-chi}
\eeq
where the lowest value corresponds to the phenomenological estimate $M_{\chi_{c1}}(2P)-M_{\chi_{c1}}(1P)\approx M_{\psi}(2S)-M_{\psi}(1S)$ while the higher one corresponds to the beginning of $t_c$-stability. This range of $t_c$-values induces an error of about 8 MeV in the meson mass determination. 
%%%%%%%%%%%%%%%%%%%%%%%%%%%%%%%
%%%%%%%%%%%%%%%%%%%%%%%%%%%%%%%%%%%%%%%
\begin{figure}[hbt]
\begin{center}
\includegraphics[width=6cm]{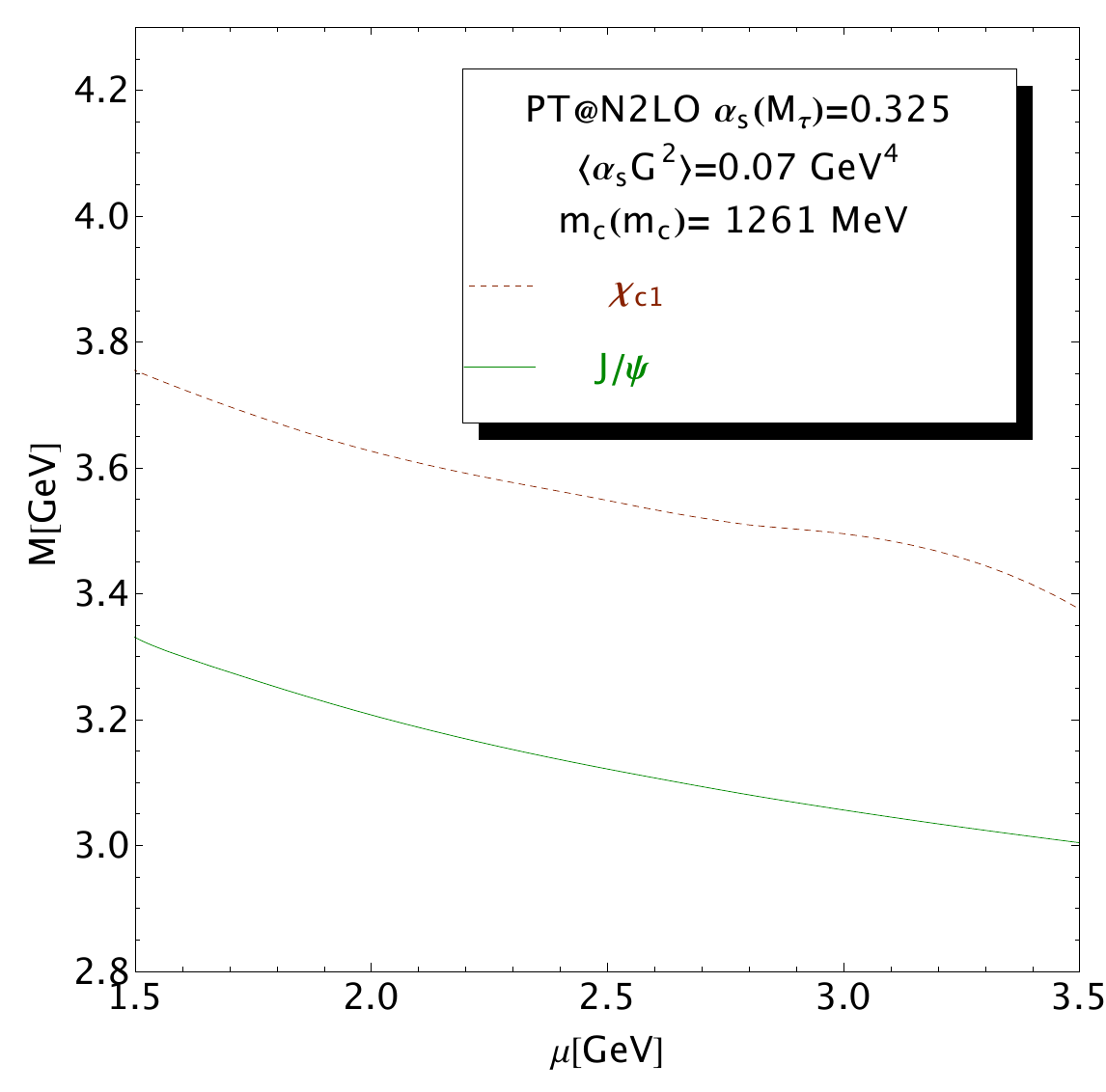}
\vspace*{-0.25cm}
\caption{\footnotesize  Behaviour of the ratio of moments ${\cal R}_{J/\psi}$ and ${\cal R}_{\chi_{c1}}$ versus $\mu$ for $t_c=20$ GeV$^2$. The inputs and the meaning of each curve are given in the legends.} 
\label{fig:muc}
\end{center}
\vspace*{-0.5cm}
\end{figure} 
%%%%%%%%%%%%%%%%%%%%%%%%%%%%%%%%%%%%%%%%%
%\vspace*{-0.25cm}
%%%%%%%%%%%%%%%%%%%%%%%%%%%%%%%%%%%%%%%%%
\subsection*{\b Subtraction point $\mu$-stability }
%%%%%%%%%%%%%%%%%%%%%%%%%%%%%%%%%%%%%%%%%
The subtraction point $\mu$ is an arbitrary parameter. It is popularly taken between 1/2 and 2 times an ``ad hoc" choice of scale. However, the physical 
observables should be not quite sensitive to $\mu$ even for a truncated PT series. In the following, like in the previous case of external (unphysical) variable, we shall fix its value by looking for a $\mu$-stability point if it exists at which the observable will be evaluated. This procedure has been used recently for improving the LSR predictions on molecules and four-quark charmonium and bottoming states\,\cite{SNSU3,SNCHI,SNCHI2,SNX,X3A}. Taking here the example of the ratios of moments, we show in Fig.\,\ref{fig:muc} their $\mu$- dependence. We notice that $ {\cal R}_{\psi}$ is a smooth decreasing function of $\mu$ while $ {\cal R}_{\chi_{c1}}$ presents a slight stability at :
\beq
\mu=(2.8\sim 2.9)~{\rm GeV},
\label{eq:muc}
\eeq
at which we shall evaluate the two ratios of moments. On can notice that at a such higher scale, one has a better convergence of the $\alpha_s(\mu)$ PT series. 
%This value of $\mu$ co\"\i ncides with the one often used choice of 3 GeV in \cite{KUHN} and may justify a such (ad hoc) choice. 
%%%%%%%%%%%%%%%%%%%%%%%%%%%%%%%%%%%%%%%
\begin{figure}[hbt]
\begin{center}
%\hspace*{-6cm}{\bf a)}\\
\includegraphics[width=8.cm]{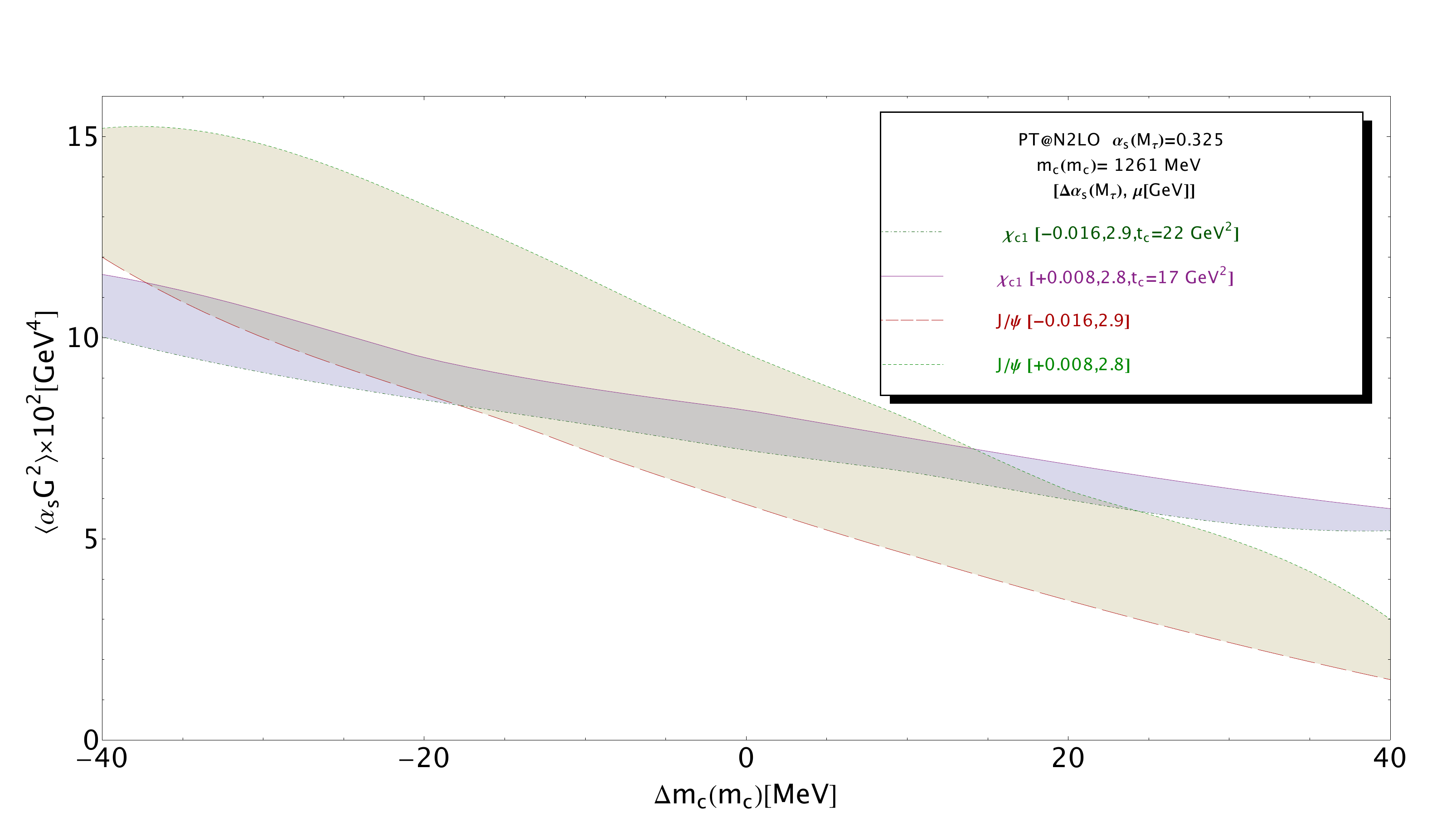}
%\hspace*{-6cm}{\bf b)}\\
%\includegraphics[width=8.cm]{mc-g29.pdf}
\vspace*{-0.5cm}
\caption{\footnotesize  Correlation between $\la\alpha_s G^2\ra$ and $\overline{m}_c(\overline{m}_c)$ for the range of $\alpha_s$ values
given in Eq.\,\ref{eq:param} and for $\mu$ given in Eq.\,\ref{eq:mub}.} 
\label{fig:mc-g2}
\end{center}
\vspace*{-0.5cm}
\end{figure} 
%%%%%%%%%%%%%%%%%%%%%%%%%%%%%%%%%%%%%%%%%
%%%%%%%%%%%%%%%%%%%%%%%%%%%%%%%%%%%%%%%%%
\subsection*{\b Correlations of the QCD parameters}
%%%%%%%%%%%%%%%%%%%%%%%%%%%%%%%%%%%%%%%%%
Once fixed these preliminaries, we are now ready to  study the correlation between $\alpha_s$, the gluon condensate $\la \alpha_s G^2\ra$,  and the $c$-quark running masses $\overline{m}_{c}(\overline{m}_{c})$.  In so doing we request that the $ \sqrt{\cal R}_{J/\psi}$ sum rule reproduces within (2-3) MeV accuracy the experimental measurement, while the $\chi_{c1}$ mass is reproduced within (8--10) MeV which is the error induced by the choice of $t_c$ in Eq.\,\ref{eq:tc-chi}. The results of the analysis are obtained at the $\tau$-stability points which are about 1.1  (resp. 0.6) GeV$^{-2}$ for the  ${J/\psi}$ (resp. $\chi_{c1}$) channels. They are shown in Fig.\,\ref{fig:mc-g2} for the two values of $\mu$ given in Eq.\,\ref{eq:muc}. One can notice that, $\la \alpha_s G^2\ra$ decreases smoother from the $\chi_{c1}$  (grey region) than from  the $J/\psi$ sum rule when $\overline{m}_{c}$ increases. In the $J/\psi$ sum rule, it moves from 0.15 to 0.02 GeV$^4$ for $\overline{m}_{c}(\overline{m}_{c})$ varying from 1221 to 1301 MeV. This feature may explain the apparent discrepancy of the results reviewed in the introduction from this channel.  

One should notice that the results from  the $J/\psi$ sum rules  are quite sensitive to the choice of the subtraction point (no $\mu$-stability) which then does not permit accurate determinations of $\la \alpha_s G^2\ra$ and $\overline{m}_{c}(\overline{m}_{c})$. Some accurate results reported in the literature for an ``ad hoc " choice of $\mu$ may be largely affected by the $\mu$ variation. 

One can also see  from  Fig.\,\ref{fig:mc-g2} that within the alone $J/\psi$ sum rule the values of $\la \alpha_s G^2\ra$ and $\overline{m}_{c}(\overline{m}_{c})$ cannot be strongly constrained\,\footnote{Similar relations from vector moments have been obtained\,\cite{IOFFEa,IOFFEb} while the ones between $\alpha_s$ and $\overline{m}_{c}$ have been studied in\,\cite{DEHNADIa,DEHNADIb}.}. Once the constraint from the $\chi_{c1}$ sum rule is introduced, one obtains a much better selection. Taking as a conservative result the range covered by the change of $\mu$ in Eq.\,\ref{eq:muc}, one deduces:
\bea
\la\alpha_s G^2\ra &=& (8.5\pm 3.0)\times 10^{-2}~{\rm GeV^4},\nnb\\
\overline{m}_{c}(\overline{m}_{c})&=&(1256\pm 30)~{\rm MeV}.
\label{eq:respsi}
\eea
We improve this determination by including the N3LO PT\,\cite{N3LO} corrections and NLO $\la \alpha_s G^2\ra$ gluon condensate (using the parametrization in \,\cite{IOFFEa,IOFFEb}) contributions \,\cite{BAIKOV}. The effects of these quantities on $\sqrt{{\cal R}_{J/\psi}}$ and $\sqrt{{\cal R}_{\chi_{c1}}}$ is about $(1\sim 2)$ MeV at the optimization scales which  induces a  negligible change such that the results quoted in Eq.\,\ref{eq:respsi} remain the same @N3LO PT and @NLO gluon condensate approximations. 
This value of  $\la \alpha_s G^2\ra$ is in good agreement with
the one $ (7.5\pm 2.0)\times 10^{-2}~{\rm GeV^4}$ from our previous analysis of the charmonium Laplace su rules using resummed PT series\,\cite{SNcb3} indicating the self-consistency of the results. However, these results do not  favor lower ones  quoted in Table\,\ref{tab:g2}.  Taking the weighted average of different sum rule  determinations given in Table\,\ref{tab:g2} with the new result in Eq.\,\ref{eq:respsi}, we obtain the {\it sum rule average}:
\beq
\la\alpha_s G^2\ra\vert_{\rm average} = (6.30\pm 0.45)\times 10^{-2}~{\rm GeV^4},
\label{eq:g2average}
\eeq
where the error may be optimistic but comparable with the one of the most precise predictions given in Table\,\ref{tab:g2}. These results agree within the errors within our recent estimates of  $\la \alpha_s G^2\ra$ and $\overline{m}_{c}(\overline{m}_{c})$\,\cite{SNcb1,SNcb2,SNcb3} 
obtained from the moments and their ratios subtracted at finite $Q^2=n\times 4m_c^2$ with $n=0,1,2.$  and from the heavy quark mass-splittings\,\cite{SNHeavy,SNHeavy2}. Hereafter, we shall use the  value of  $\la \alpha_s G^2\ra$ in Eq.\,\ref{eq:g2average}.
%%%%%%%%%%%%%%%%%%%%%%%%%%%%%%%%%%%%%%%
\begin{figure}[hbt]
\begin{center}
\includegraphics[width=10cm]{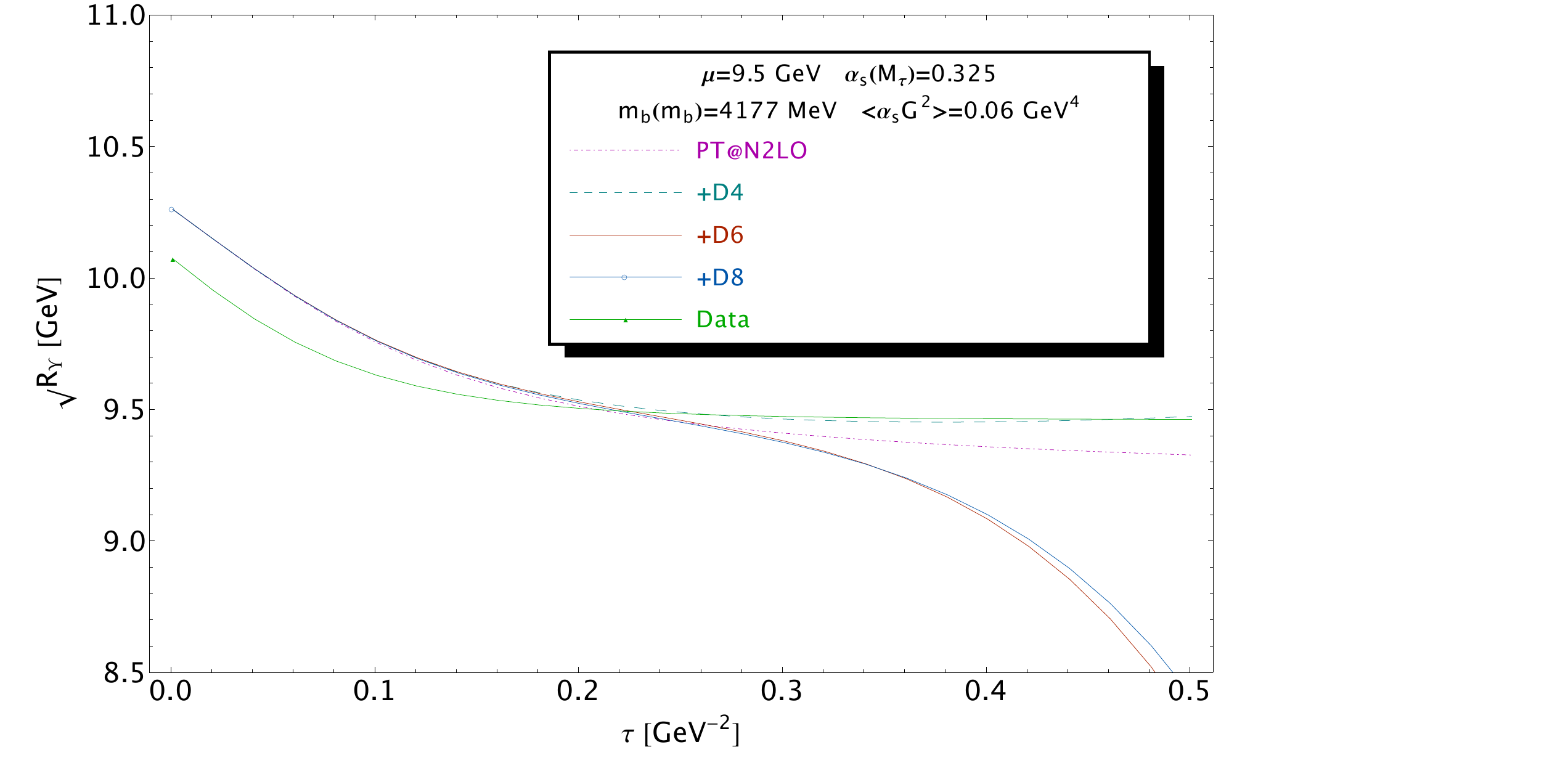}
\vspace*{-0.5cm}
\caption{\footnotesize  Behaviour of the ratio of moments ${\cal R}_{\Upsilon}$ versus $\tau$ in GeV$^{-2}$ for different truncation of the OPE. The input and the meaning of each curve are given in the legend.} 
\label{fig:upsilon-tau}
\end{center}
\vspace*{-0.8cm}
\end{figure} 
%%%%%%%%%%%%%%%%%%%%%%%%%%%%%%%%%%%%%%%%%%%%%%%%%%%
%%%%%%%%%%%%%%%%%%%%%%%%%%%%%%%%%%%%%%%
\begin{figure}[hbt]
\begin{center}
\includegraphics[width=10.5cm]{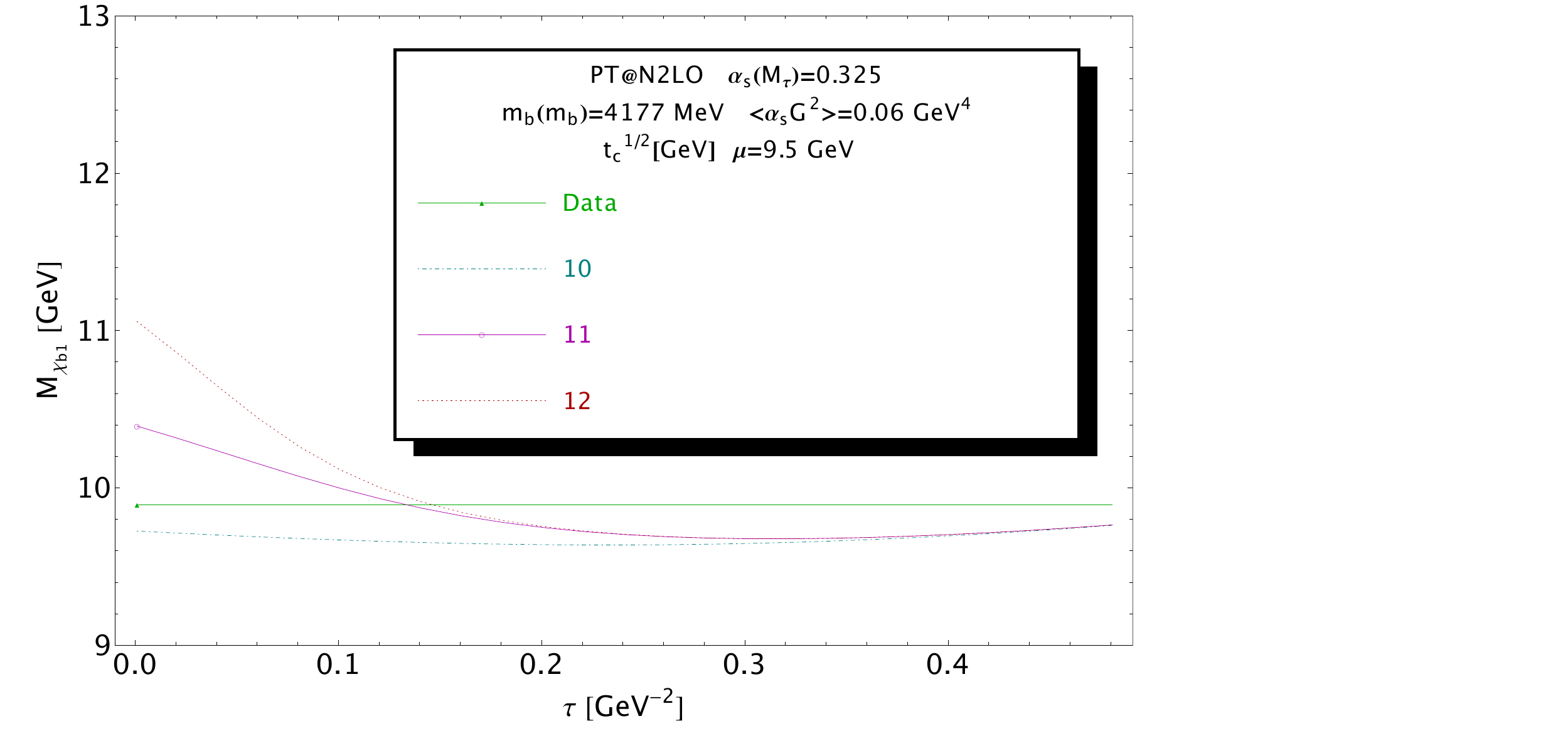}
\vspace*{-0.5cm}
\caption{\footnotesize  Behaviour of the ratio of moments ${\cal R}_{\chi_{b1}}$ versus $\tau$ for different values of $t_c$. The input and the meaning of each curve are given in the legend.} 
\label{fig:chi1b-tc}
\end{center}
\vspace*{-0.7cm}
\end{figure} 
%%%%%%%%%%%%%%%%%%%%%%%%%%%%%%%%%%%%%%%%%
%%%%%%%%%%%%%%%%%%%%%%%%%%%%%%%%%%%%%%%%%
\section{Bottomium Ratios of Moments $ {\cal R}_{\Upsilon(\chi_{b1})}$ }
%%%%%%%%%%%%%%%%%%%%%%%%%%%%%%%%%%%%%%%%%
\subsection*{\b $\tau$ and $t_c$-stabilities and test of convergences}
%%%%%%%%%%%%%%%%%%%%%%%%%%%%%%%%%%%%%%%%
The analysis is very similar to the previous $J/\psi$ sum rule. The relative perturbative and non-perturbative contributions are very similar to the curves in Figs.\,\ref{fig:pertc} to \ref{fig:npertc}.
 We use the value:
 $ \mu=9.5$ GeV  which we shall justify later on. 
 However, it is informative to show in Fig.\ref{fig:upsilon-tau} the $\tau$-behaviour of ${\cal R}_{\Upsilon}$ for different truncation of the OPE where $\tau$-stability is obtained at $\tau\simeq 0.22$ GeV$^{-2}$. In Fig.\,\ref{fig:chi1b-tc}, we show the $\tau$-behaviour of ${\cal R}_{\chi_{b1}}$ for different values of $t_c$ from which we deduce a stability at $\tau\simeq 0.28$ GeV$^{-2}$ and $t_c$-stability which we shall take to be $\sqrt{t_c}\simeq 11$ GeV.  A much better convergence of the $\alpha_s$ series is observed as the sum rule is evaluated at a higher scale $\mu$. The OPE converges also faster as $\tau$ is smaller here. 
%%%%%%%%%%%%%%%%%%%%%%%%%%%%%%%%%%%%%%%%%
\subsection*{\b $\mu$-stability}
\begin{figure}[hbt]
\vspace*{-0.5cm}
\begin{center}
\includegraphics[width=7.5cm]{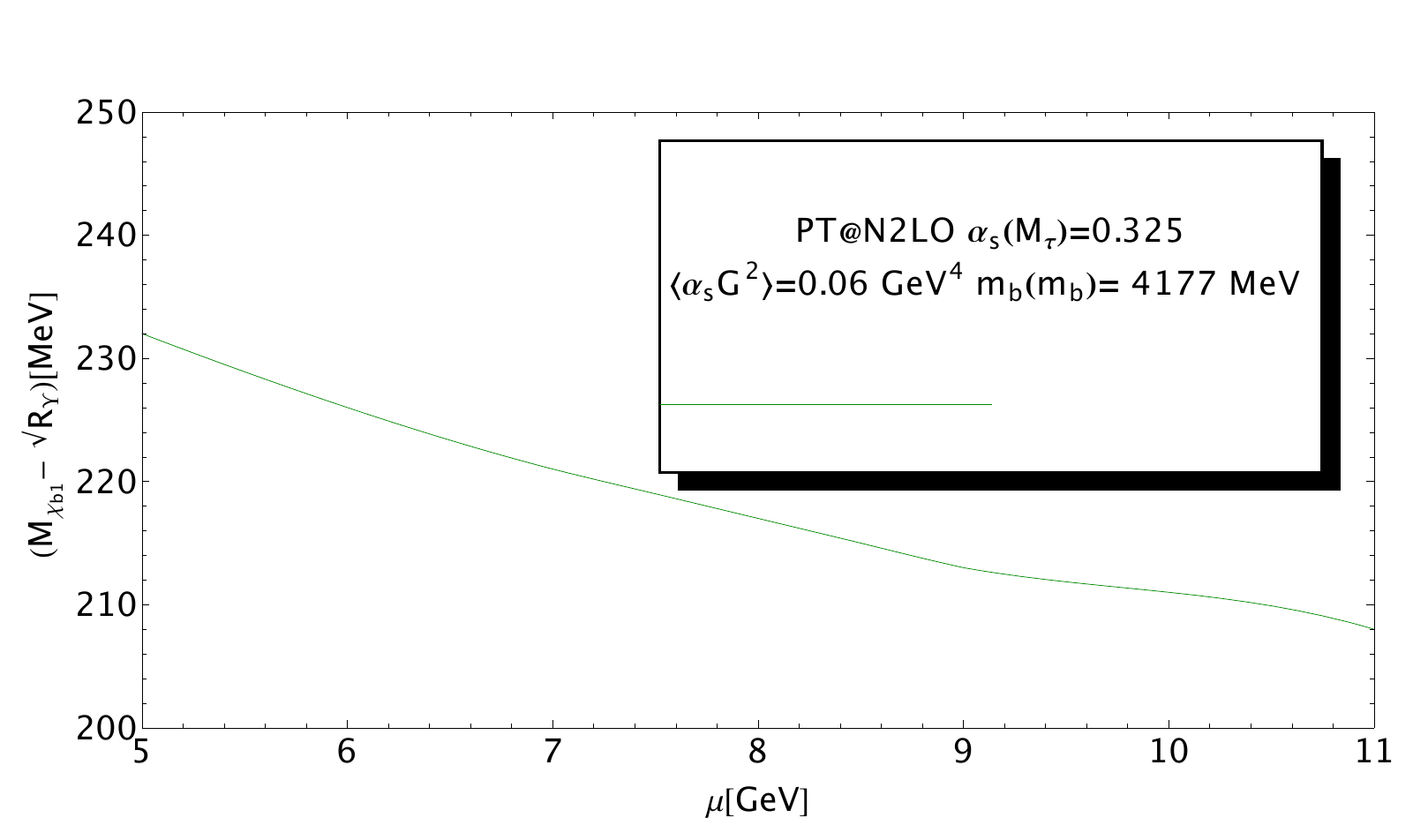}
\vspace*{-0.5cm}
\caption{\footnotesize  Behaviour of $M_{\chi_{b1}}-\sqrt{{\cal R}_{\Upsilon}}$ versus $\mu$.} 
\label{fig:delta}
\end{center}
\vspace*{-1cm}
\end{figure}

%%%%%%%%%%%%%%%%%%%%%%%%%%%%%%%%%%%%%%%%
The two sum rules are smooth decreasing functions of $\mu$ but does not show $\mu$-stability. Instead, their difference presents $\mu$-stability at:
\beq
\mu\simeq (9\sim 10)~{\rm GeV},
\label{eq:mub}
\eeq
 as shown in Fig.\ref{fig:delta} at which we choose to evaluate the two sum rules. 
%%%%%%%%%%%%%%%%%%%%%%%%%%%%%%%%%%%%%%%
%%%%%%%%%%%%%%%%%%%%%%%%%%%%%%%%%%%%%%%%% 
%%%%%%%%%%%%%%%%%%%%%%%%%%%%%%%%%%%%%%%
\begin{figure}[hbt]
\begin{center}
\includegraphics[width=7.5cm]{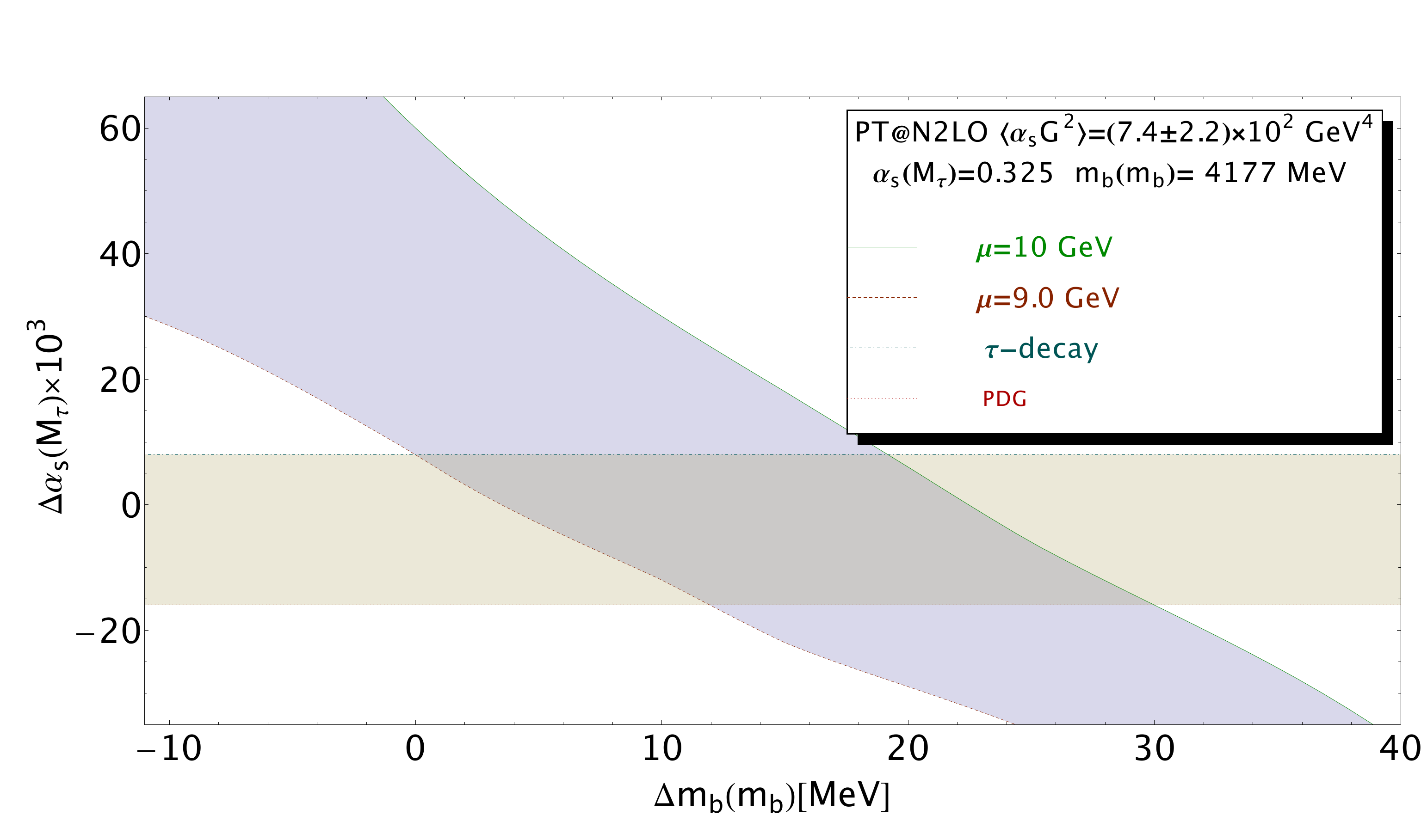}
%\vspace*{-0.7cm}
\caption{\footnotesize  Behaviour of $\Delta\alpha_s(M_\tau)$ versus $\overline{m}_{b}(\overline{m}_{b})$from the ratio of moments ${\cal R}_{\Upsilon}$ The horizontal band corresponds to the range of $\alpha_s$ value given in Eq.\,\ref{eq:param}. The input and the meaning of each curve are given in the legend.} 
\label{fig:mb-alfas}
\end{center}
\vspace*{-0.5cm}
\end{figure} 
%%%%%%%%%%%%%%%%%%%%%%%%%%%%%%%%%%%%%%%%%
%%%%%%%%%%%%%%%%%%%%%%%%%%%%%%%%%%%%%%%%%
\subsection*{\b Mass of $\chi_{b1}(1^{++})$ from $ {\cal R}_{\chi_{b1}}$}
%%%%%%%%%%%%%%%%%%%%%%%%%%%%%%%%%%%%%%%%%
Using  the previous value of the QCD parameters, we predict from the ratio of $\chi_{b1}$ moments:
\beq
M_{\chi_{b1}}\simeq 9677(26)_{t_c}(8)_{\alpha_s}(11)_{G^2}(9)_{m_b}(99)_\mu~{\rm MeV}~,
\eeq
which is (within the error) about 100 MeV lower than the experimental mass $M_{\chi_{b1}}^{exp}=9893$ MeV. The agreement between theory and experiment may be improved when more data for higher states are available or/and by including Coulombic corrections shown to be small for the vector current (see e.g\,\cite{SNcb1}) and not considered here. 
%%%%%%%%%%%%%%%%%%%%%%%%%%%%%%%%%%%%%%%%%
\subsection*{\b Correlation between $\alpha_s(\mu)$ and $\overline{m}_{b}(\overline{m}_{b})$ from  $ {\cal R}_{\Upsilon }$}
%%%%%%%%%%%%%%%%%%%%%%%%%%%%%%%%%%%%%%%%
From the previous analysis, one can notice that the $\chi_{b1}$ channel cannot help from a precise  study of the correlation between $\alpha_s$ and $\overline{m}_{b}(\overline{m}_{b})$.  We show in Fig.\,\ref{fig:mb-alfas} the result of the analysis from the $\Upsilon$ channel by requiring that the experimental value of $ \sqrt{\cal R}_{\Upsilon }$ is reproduced within $(1\sim 2)$ MeV accuracy. First, one can notice that the error due to the gluon condensate with the value given in Eq.\,\ref{eq:respsi} is negligible. 
Given the range of $\alpha_s$ quoted in Eq.\,\ref{eq:param}, one can deduce the prediction:
\beq
\overline{m}_{b}(\overline{m}_{b})=4192(15)(8)_{coul}~{\rm MeV}~,
\label{eq:resmb}
\eeq
where we have added in Eq.\,\ref{eq:resmb} an error of about 8 MeV from Coulombic corrections as estimated in \cite{SNcb3}. 
The previous result in Eq.\,\ref{eq:resmb} corresponds to:
\beq
\hspace*{-0.5cm} \alpha_s(M_\tau)=0.321(12) \lrar\alpha_s(M_Z)=0.1186(15)(3)
\label{eq:alfas-mb}
 \eeq  
given by the range in Eq.\,\ref{eq:param}. The running from $M_\tau$ to $M_Z$ due to the choice of the thresholds induces the last error (3).
%%%%%%%%%%%%%%%%%%%%%%%%%%%%%%%%%%%%%%%%%
\subsection*{\b Updated average value of $\overline{m}_{b}(\overline{m}_{b})$ from QSSR}
%%%%%%%%%%%%%%%%%%%%%%%%%%%%%%%%%%%%%%%%
The result in Eq.\,\ref{eq:resmb} is consistent with the ones from LSR with RG resummed PT expressions\,\cite{SNcb3}:
\beq
\overline{m}_{b}(\overline{m}_{b})=4212(32)~{\rm MeV}~,
\label{eq:mblsr}
\eeq 
and the average of the ones from moments sum rules quoted in Eq.\,\ref{eq:param} and updated in\,\cite{SNmass18}: 
\beq
\overline{m}_{b}(\overline{m}_{b})=4188(8)~{\rm MeV}~.
\label{eq:mbmom}
\eeq 
  Taking the average of the LSR and updated moments determinations, we obtain the final estimate:
\beq
\overline{m}_{b}(\overline{m}_{b})\vert_{\rm average}=(4190\pm 8)~{\rm MeV}~,
\label{eq:mbaverage}
\eeq
where the errors come from the most precise determination. 
%Due to the large errors induced by the subtraction scale as shown in Fig\,\ref{fig:mb-alfas}, one cannot accurately extract from the the value of $\alpha_s$ given the present value of $\overline{m}_{b}$.
%%%%%%%%%%%%%%%%%%%%%%%%%%%%%%%%%%%%%%%
\begin{figure}[hbt]
\begin{center}
\includegraphics[width=7.8cm]{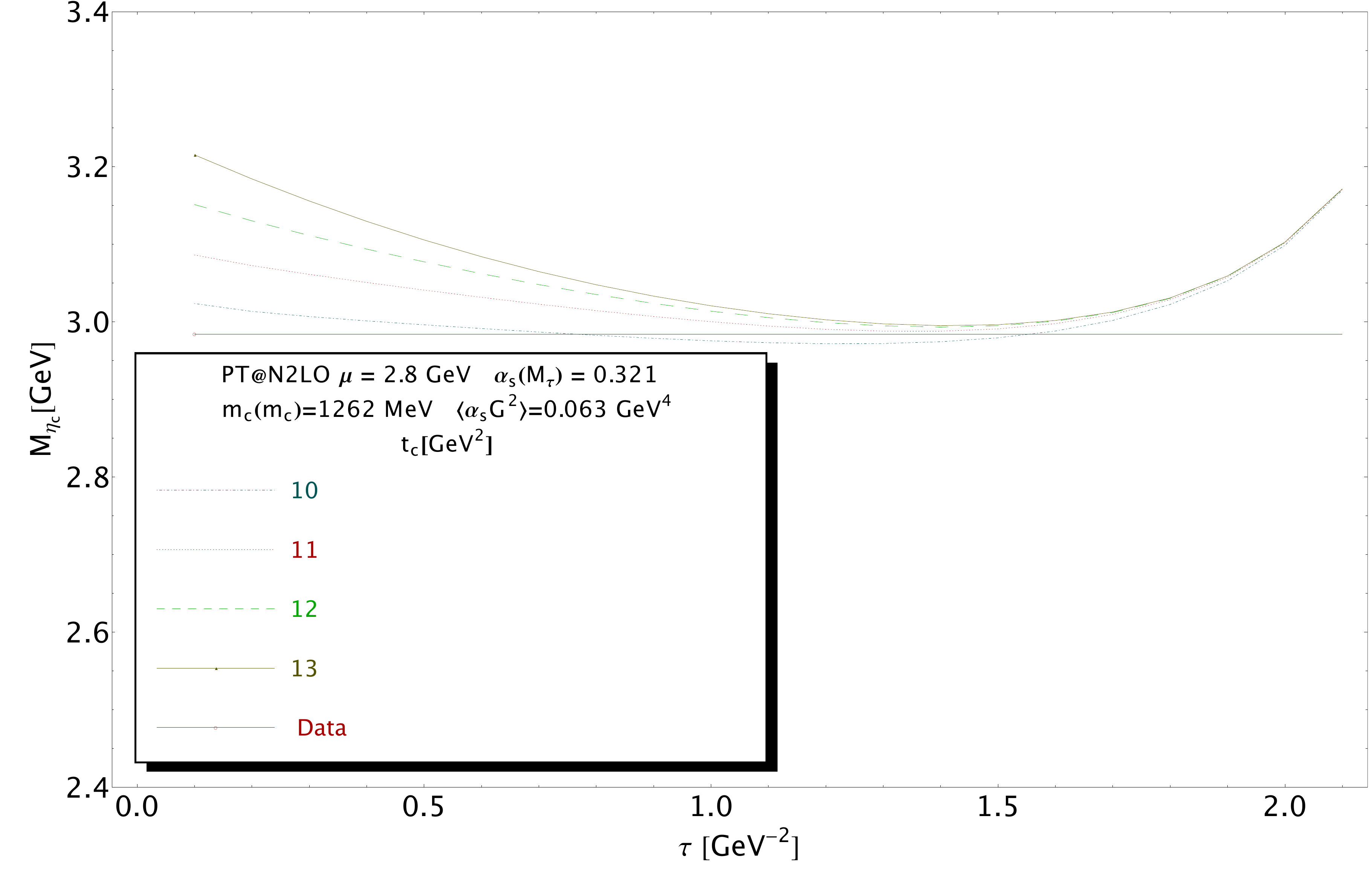}
\vspace*{-0.5cm}
\caption{\footnotesize  Behaviour of $M_{\eta_c}$ versus $\tau$ for different values of $t_c$.} 
\label{fig:etac-tau}
\end{center}
\vspace*{-0.5cm}
\end{figure} 
%%%%%%%%%%%%%%%%%%%%%%%%%%%%%%%%%%%%%%%%% 
 %%%%%%%%%%%%%%%%%%%%%%%%%%%%%%%%%%%%%%%
\begin{figure}[hbt]
\begin{center}
\includegraphics[width=7.8cm]{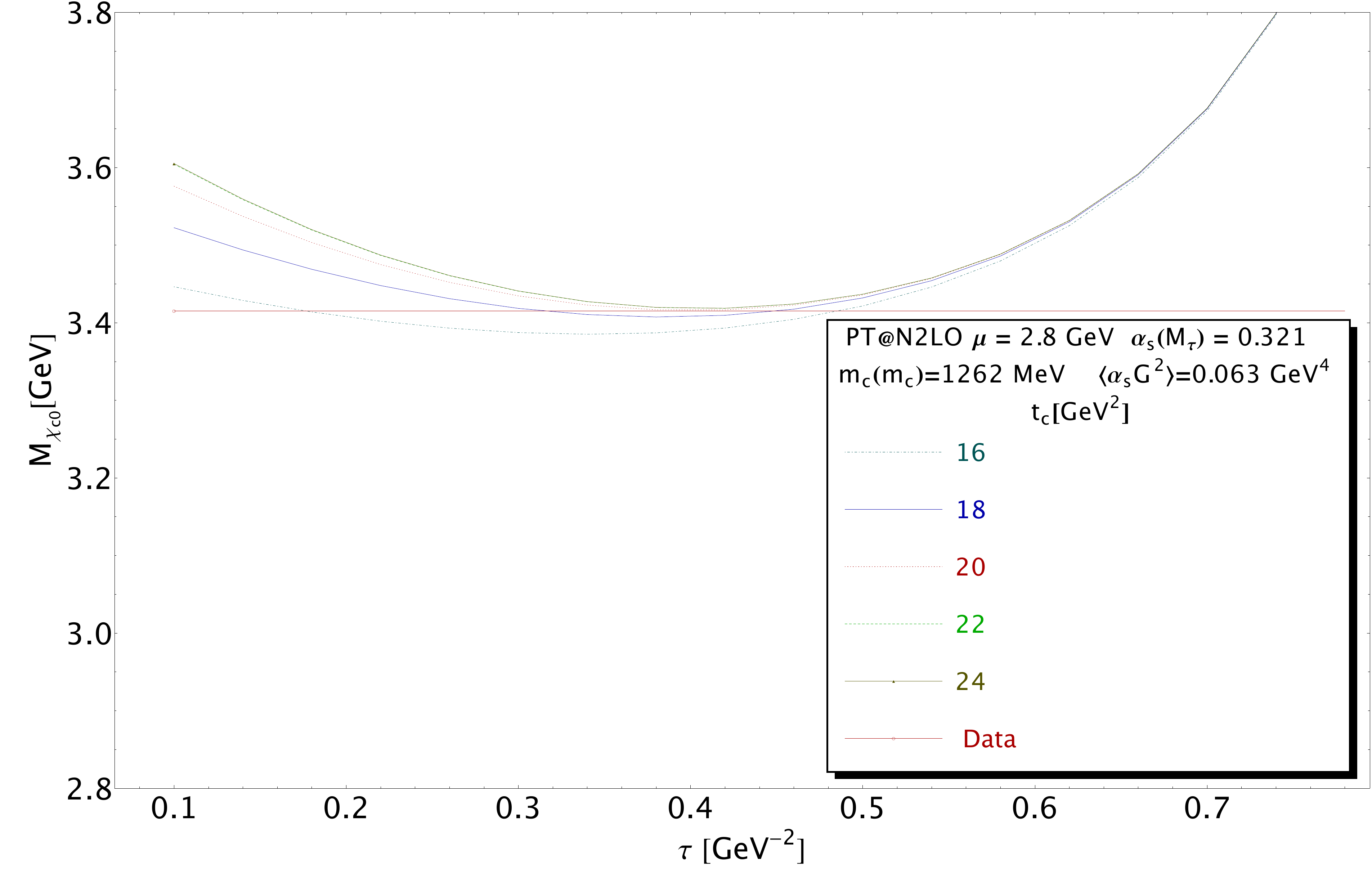}
\vspace*{-0.5cm}
\caption{\footnotesize  Behaviour of $M_{\chi_{c0}}$ versus $\tau$ for different values of $t_c$.} 
\label{fig:chic-tau}
\end{center}
\vspace*{-0.7cm}
\end{figure} 
%%%%%%%%%%%%%%%%%%%%%%%%%%%%%%%%%%%%%%%%% 
%%%%%%%%%%%%%%%%%%%%%%%%%%%%%%%%%%%%%%%%%
\section{(Pseudo)scalar charmonium}
%%%%%%%%%%%%%%%%%%%%%%%%%%%%%%%%%%%%%%%%%
In these channels,  we shall work with the ratio of sum rules associated to the  two-point correlator $\Psi_{P(S)}(q^2)$ defined in Eq.\,\ref{eq:2-pseudo} which is not affected by $\Psi_{P(S)}(0)$. We shall use the PT expression known @N2LO\,\cite{TEUBNER,HOANGa,CHET1,CHET2,CHET3}, the contribution of the gluon condensates of dimension 4 and 6 to LO\,\cite{NIKOLa,NIKOLb}. 
%%%%%%%%%%%%%%%%%%%%%%%%%%%%%%%%%%%%%%%%%
\subsection*{\b $\eta_c$ and $\chi_{c0}$ masses}
%%%%%%%%%%%%%%%%%%%%%%%%%%%%%%%%%%%%%%%%%
  The $\eta_c$ sum rule shows a smooth decreasing function of $\mu$ but does not present a $\mu$-stabiity. Then, we choose the value of $\mu$ given in Eq.\,\ref{eq:muc} for evaluating it.  We show in Fig.\,\ref{fig:etac-tau} the $\tau$-behaviour of the $\eta_c$-mass
 for different values of $t_c$ which we take from 10 GeV$^2$ [around the mass squared of the $\eta_c(2P)$ and  $\eta_c(3P)$]  until 13 GeV$^2$ ($t_c$-stability) . Similar analysis is done for the $\chi_{c0}$ associated to the scalar current $\bar Q(i)Q$ which is shown in Fig.\ref{fig:chic-tau}, where we take $t_c\simeq (16\sim 24)$ GeV$^2$.
  Using the averaged values of $\la \alpha_s G^2\ra$ and  $\overline{m}_{c}(\overline{m}_{c})$ in Eqs.\,\ref{eq:g2average} and \ref{eq:mcaverage}, we deduce the optimal result in units of MeV:
 \bea
 M_{\eta_c}&=&2979(5)_\mu(11)_{t_c}(11)_{\alpha_s}(30)_{m_c}(10)_{G^2}~,\nnb\\
  M_{\chi_{c0}}&=&3411(1)_\mu(17)_{t_c}(26)_{\alpha_s}(30)_{m_c}(20)_{G^2}~,
 \label{eq:chic}
 \label{eq:etac}
 \eea
 in good agreement within the errors with the experimental masses:  $M_{\eta_c}=2984$ MeV and $M_{\chi_{c0}}$=3415 MeV but not enough accurate for extracting with precision the QCD parameters. 
 %%%%%%%%%%%%%%%%%%%%%%%%%%%%%%%%%%%%%%%%%
\subsection*{\b Correlation between $\overline{m}_{c}(\overline{m}_{c})$ and $\la \alpha_s G^2\ra$}
\begin{figure}[hbt]
\begin{center}
\includegraphics[width=8.5cm]{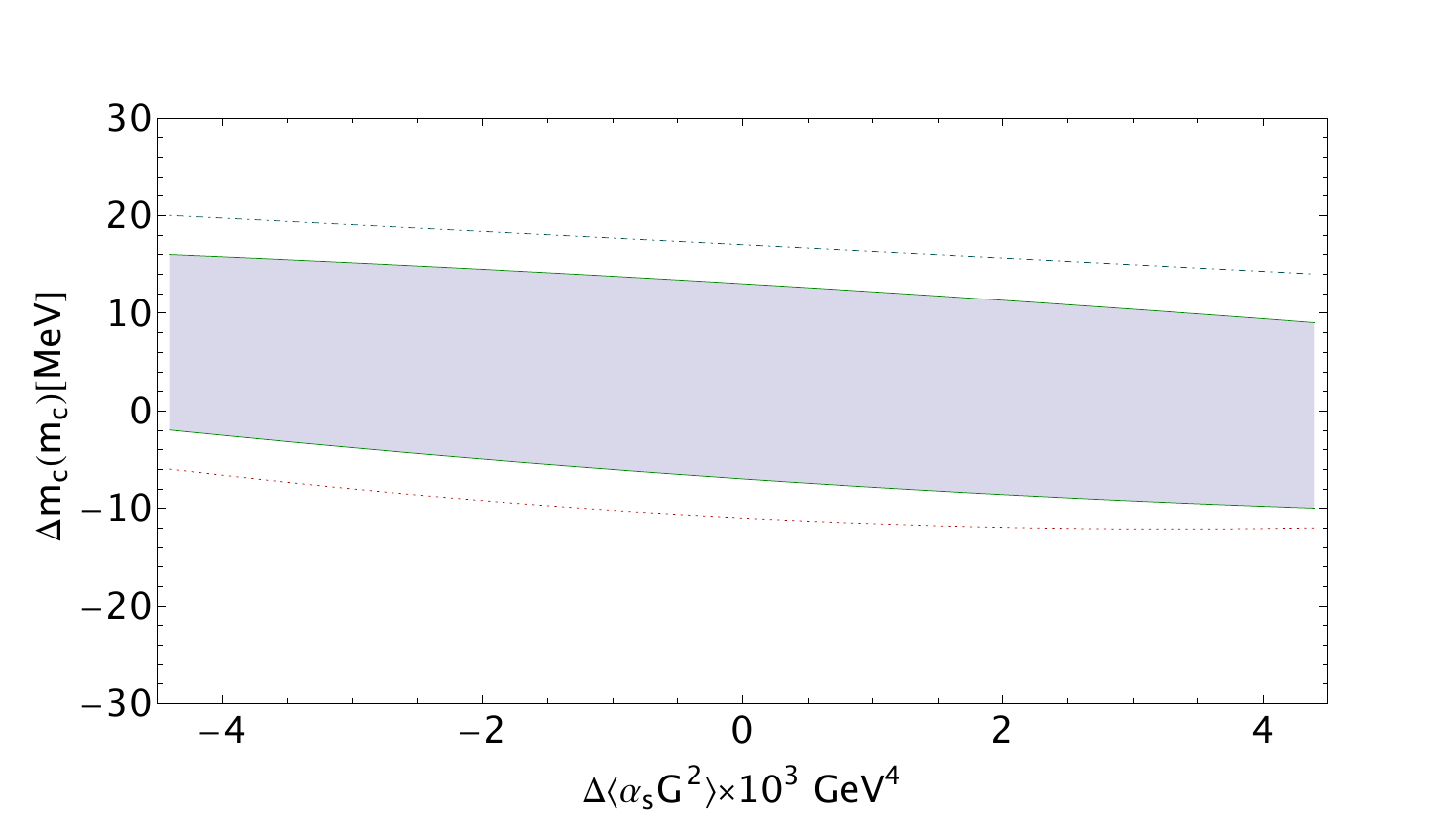}
\vspace*{-0.5cm}
\caption{\footnotesize  Behaviour of $\Delta \overline{m}_{c}(\overline{m}_{c})$ versus $\Delta\la \alpha_s G^2\ra$ from $M_{\eta_c}$. The dashed region corresponds to $\Delta \alpha_s=0$ and for different values of $t_c\simeq 10\sim 13$ GeV$^2$. The two extremal lines correspond to $\Delta\alpha_s\pm 12$. We use the range of $\alpha_s$  in Eq.\,\ref{eq:param} and of $\la \alpha_s G^2\ra$ in Eq.\,\ref{eq:g2average}. }
\label{fig:mc-g2-etac}
\end{center}
\vspace*{-0.5cm}
\end{figure}

%%%%%%%%%%%%%%%%%%%%%%%%%%%%%%%%%%%%%%%%%
We study the correlation between $\overline{m}_{c}(\overline{m}_{c})$ and $\la \alpha_s G^2\ra$ by requiring that the sum rules reproduce the masses of the ${\eta_c}$ and ${\chi_{c0}}$ within the error induced by the choice of $t_c$ repsectively 11 and 17 MeV. We show the result of the analysis in Fig.\,\ref{fig:mc-g2-etac} keeping only the strongest constraint from $M_{\eta_c}$. We deduce:
\beq
\overline{m}_{c}(\overline{m}_{c})=1266(16) ~{\rm MeV}~,
\label{eq:mcetac}
\eeq
 %%%%%%%%%%%%%%%%%%%%%%%%%%%%%%%%%%%%%%%
% \vspace*{-0.5cm}
%%%%%%%%%%%%%%%%%%%%%%%%%%%%%%%%%%%%%%%%%
in good agreement with the one in \,\cite{ZYAB} from pseudoscalar moments. 
%%%%%%%%%%%%%%%%%%%%%%%%%%%%%%%%%%%%%%%%%
\subsection*{\b Updated average value of $\overline{m}_{c}(\overline{m}_{c})$ from QSSR}
%%%%%%%%%%%%%%%%%%%%%%%%%%%%%%%%%%%%%%%%
We combine our determinations in Eqs.\,\ref{eq:respsi} and \ref{eq:mcetac} 
%The value of $\overline{m}_{c}(\overline{m}_{c})$ which we have derived from the value $\overline{m}_{c}(\mu)$ with $\mu$ given in Eq.\,\ref{eq:muc} at which the sum rule has been evaluated also agrees with the average of
with the updated determination\,\cite{SNmass18}:
 \beq
\overline{m}_{c}(\overline{m}_{c})\vert_{\rm average}=(1264\pm 6)~{\rm MeV},
\label{eq:mcmom1}
 \eeq
 of two ones\,\cite{SNcb1,SNcb2} from vector moments sum rules quoted in Eq.\,\ref{eq:param}.
 As a final result, we quote the updated average from exponential and moment sum rules from a global fit of the quarkonia spectra:
 \beq
\overline{m}_{c}(\overline{m}_{c})\vert_{\rm average}=(1264\pm 6)~{\rm MeV},
\label{eq:mcaverage}
 \eeq
 which is dominated by the most precise prediction quoted in Eq.\,\ref{eq:mcmom1}. 
  It is remarkable that this value agrees with the original SVZ estimate\,\cite{SVZa,SVZb} of the euclidian mass.
 %%%%%%%%%%%%%%%%%%%%%%%%%%%%%%%%%%%%%%%%%
\section{(Pseudo)scalar bottomium}
%%%%%%%%%%%%%%%%%%%%%%%%%%%%%%%%%%%%%%%%%
%%%%%%%%%%%%%%%%%%%%%%%%%%%%%%%%%%%%%%%%%
\subsection*{\b $\eta_b$ and $\chi_{b0}$ masses}
%%%%%%%%%%%%%%%%%%%%%%%%%%%%%%%%%%%%%%%%%
 %%%%%%%%%%%%%%%%%%%%%%%%%%%%%%%%%%%%%%%
\begin{figure}[hbt]
\begin{center}
\includegraphics[width=7cm]{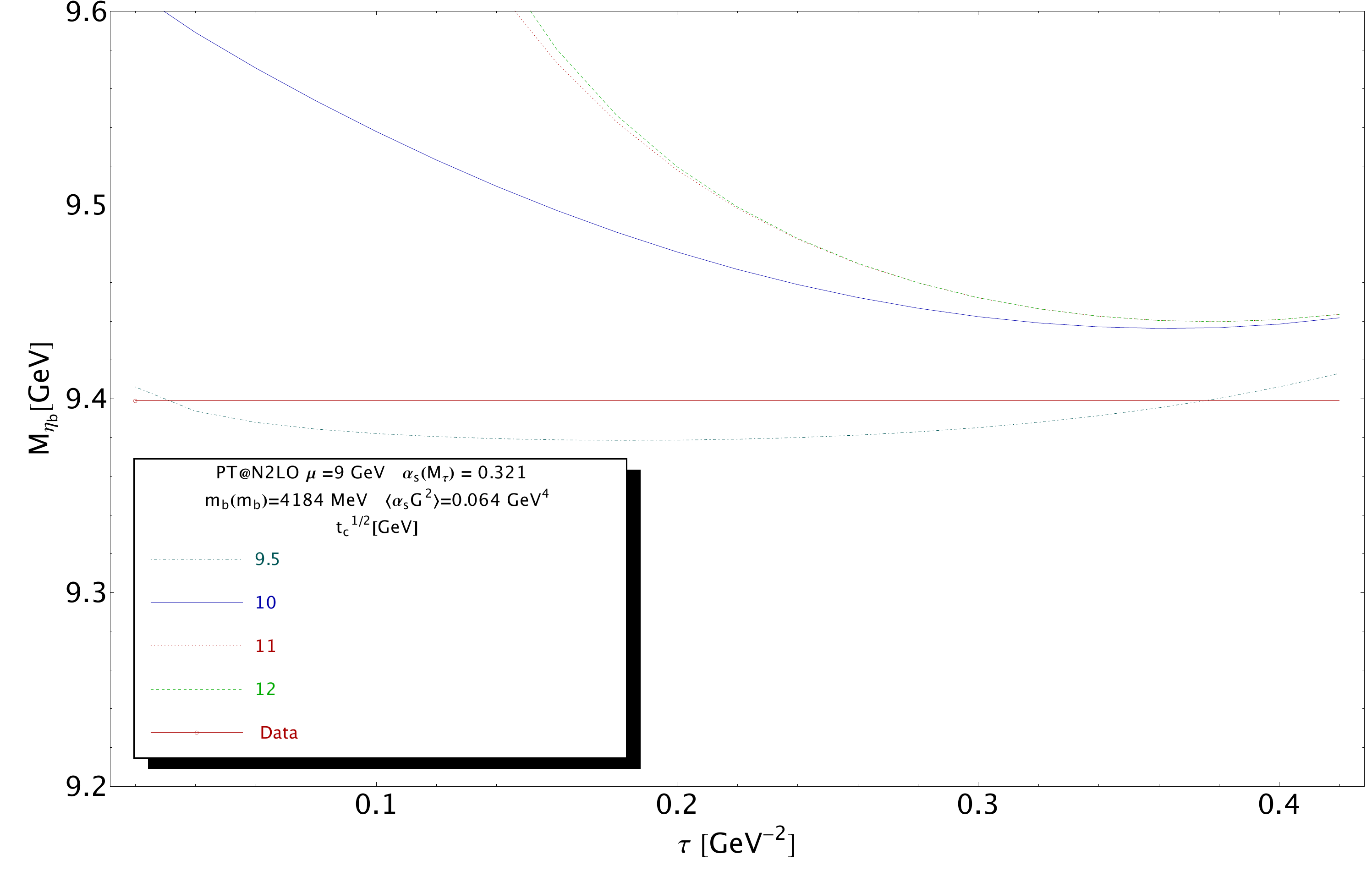}
\vspace*{-0.5cm}
\caption{\footnotesize  Behaviour of $M_{\eta_b}$ versus $\tau$ for different values of $t_c$.} 
\label{fig:etab}
\end{center}
\vspace*{-0.5cm}
\end{figure} 
%%%%%%%%%%%%%%%%%%%%%%%%%%%%%%%%%%%%%%%%% 
\begin{figure}[hbt]
\begin{center}
\includegraphics[width=7.7cm]{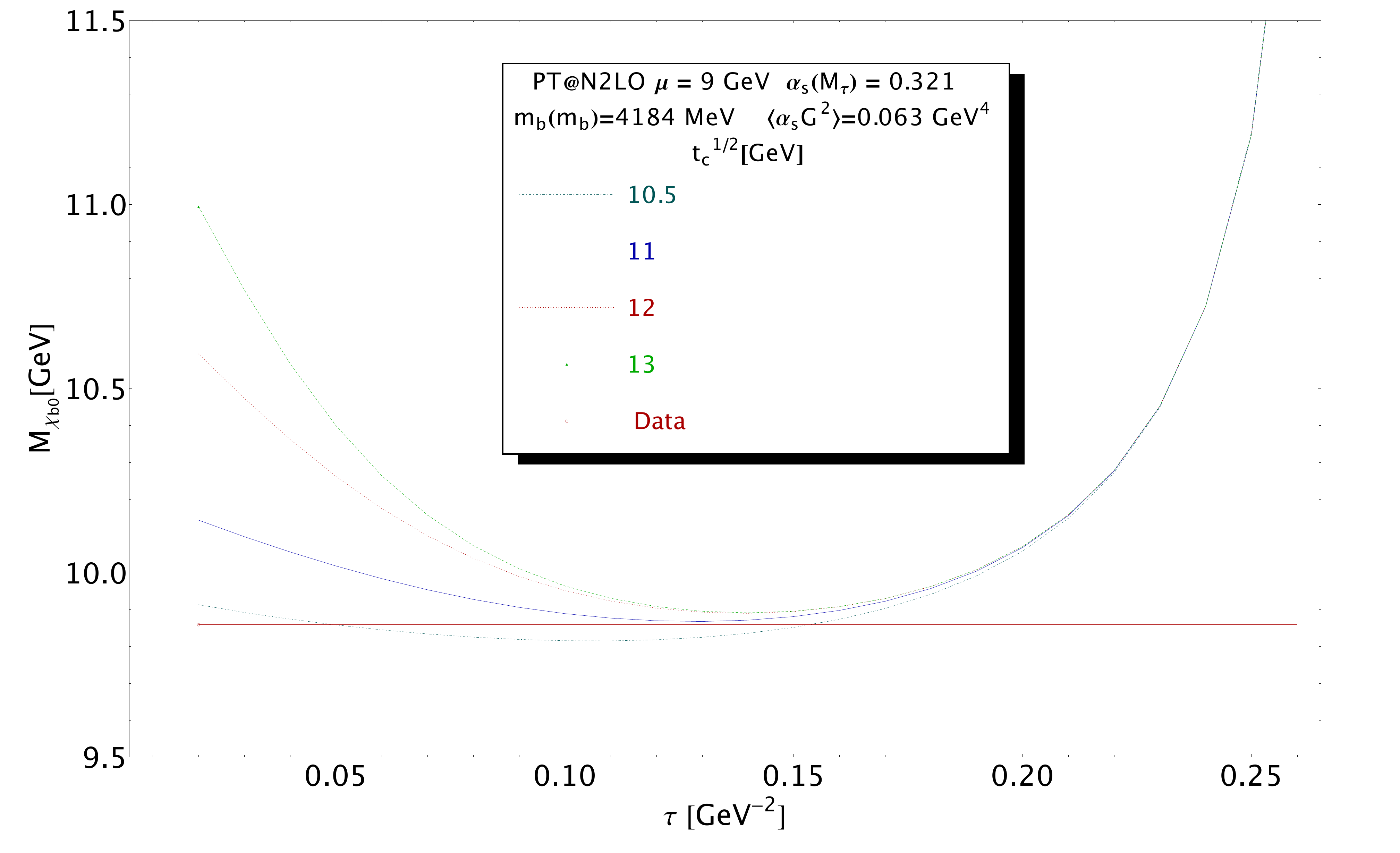}
\vspace*{-0.5cm}
\caption{\footnotesize  Behaviour of $M_{\chi_{b0}}$ versus $\tau$ for different values of $t_c$.} 
\label{fig:chib}
\end{center}
\vspace*{-0.25cm}
\end{figure} 
  %%%%%%%%%%%%%%%%%%%%%%%%%%%%%%%%%%%%%%%
The masses of the $\eta_b(0^{-+})$ and $\chi_{b0}(0^{++})$ are extracted in a similar way using the value of $\mu$ in Eq.\,\ref{eq:mub}
and the parameters in Eqs.\,\ref{eq:g2average} and \ref{eq:mbaverage}. We take the range $\sqrt{t_c}=(9.5\sim 12)$  [resp. $(10.5 \sim 13)$] GeV for the $\eta_b$ [resp. $\chi_{b0}$] channels, as shown in Figs\,\ref{fig:etab} and \,\ref{fig:chib} from which we deduce in units of MeV:
\bea
M_{\eta_b}&=&9394(16)_\mu(30)_{t_c}(7)_{\alpha_s}(16)_{m_b}(8)_{G^2}~,\nnb\\
M_{\chi_{b0}}&=&9844(7)_\mu(35)_{t_c}(6)_{\alpha_s}(17)_{m_b}(29)_{G^2}~,
\label{eq:mchib}
\label{eq:metab}
\eea
 %%%%%%%%%%%%%%%%%%%%%%%%%%%%%%%%%%%%%%% %%%@%%
in good agreement with the data  $M_{\eta_b}=9399$ MeV and $M_{\chi_{b0}}=9859$ MeV.
 %%%%%%%%%%%%%%%%%%%%%%%%%%%%%%%%%%%%%%%%%%%%%%%%%%
\subsection*{\b Correlation between $\overline{m}_{b}(\overline{m}_{b})$ and $\la \alpha_s G^2\ra$}
%%%%%%%%%%%%%%%%%%%%%%%%%%%%%%%%%%%%%%%%%%%%%%%%%%
The analysis done for charmonium is repeated here where we request that the sum rule reproduces the $\eta_b$ and  $\chi_{b0}$ masses with the error induced by the choice of $t_c$. Unfortunately, this constraint is too weak and leads to $\overline{m}_{b}(\overline{m}_{b})$ with an accuracy of about 40 MeV which is less interesting than the estimate from the vector channel in Eq.\,\ref{eq:resmb}. 
 %%%%%%%%%%%%%%%%%%%%%%%%%%%%%%%%%%%%%%%%%%%%
\section{$\alpha_s$ and $\la \alpha_s G^2\ra$ from $M_{\chi_{0c(0b)}}-M_{\eta_{c(b)}}$}
\begin{figure}[hbt]
\vspace*{-0.5cm}
\begin{center}
\includegraphics[width=8.cm]{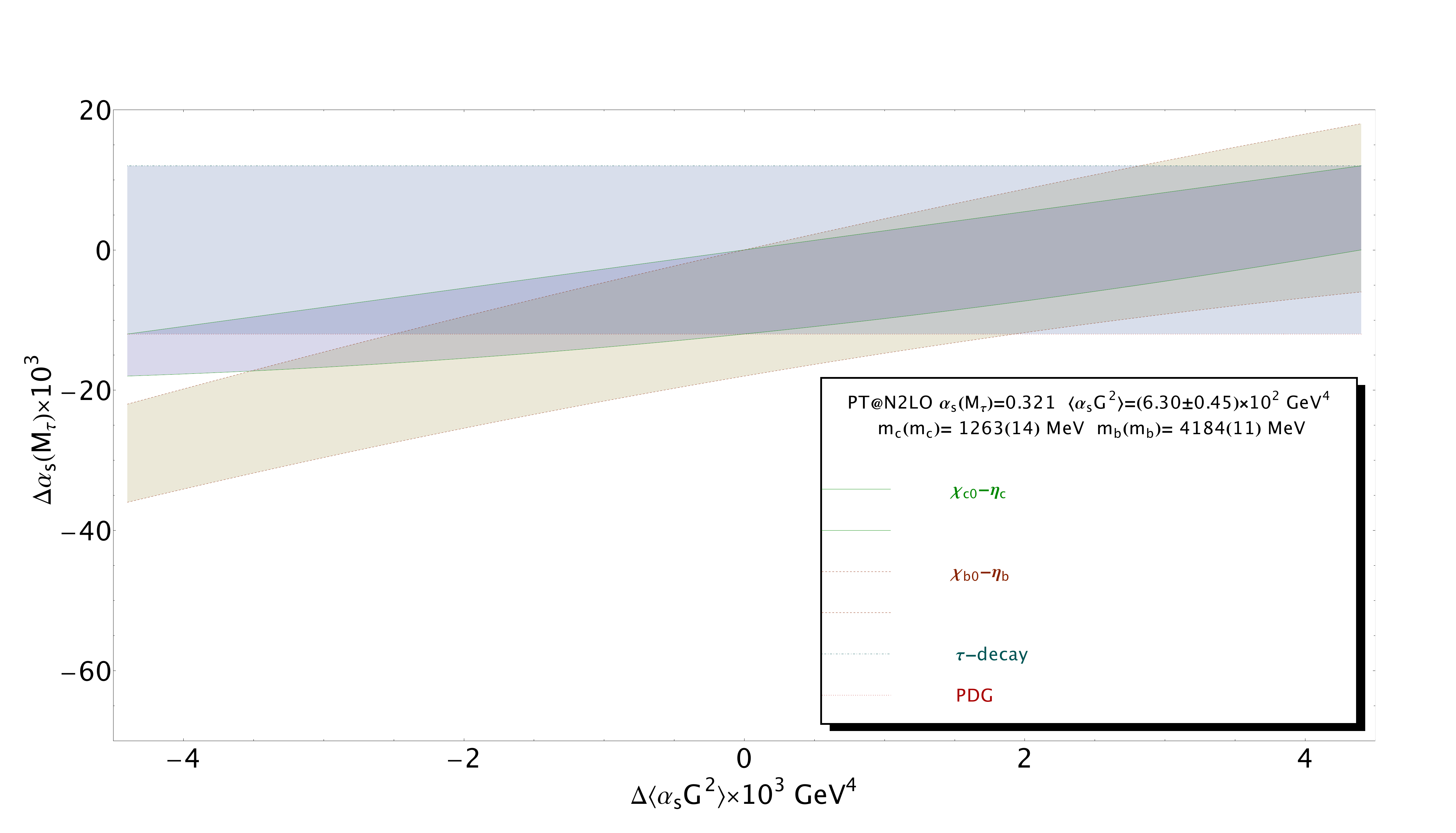}
\vspace*{-0.25cm}
\caption{\footnotesize  Correlation between $\alpha_s$ and $\la \alpha_s G^2\ra$ by requiring that the sum rules reproduce the (pseudo)scalar mass-splittings.} 
\label{fig:alfas-g2}
\end{center}
%\vspace*{-0.75cm}
\end{figure}

\begin{figure}[hbt]
%\vspace*{-0.5cm}
\begin{center}
\includegraphics[width=8.cm]{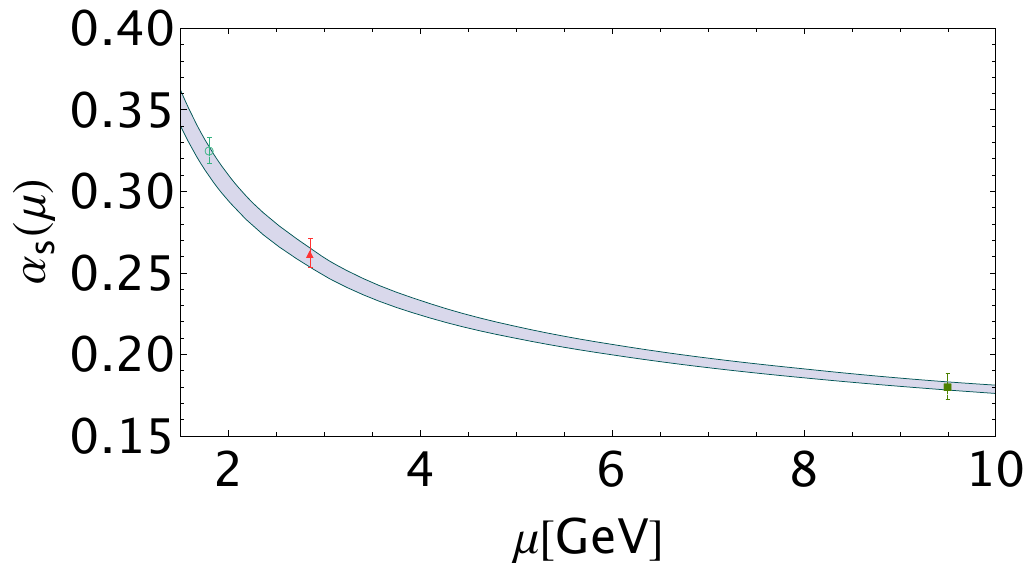}
\vspace*{-0.25cm}
\caption{\footnotesize  Comparison with the running of the world average $\alpha_s(M_Z)=0.1181(11)$\,\cite{BETHKEa,PDG} of our predictions at three different scales: $M_\tau$  for the original low moment $\tau$-decay width\,\cite{SNTAU} (open circle),  2.85 GeV for $M_{\chi_{c0}}-M_{\eta_c}$ (full triangle) and  9.5 GeV for $M_{\chi_{b0}}-M_{\eta_b}$ (full square)\,\cite{SN18}.}
\label{fig:alfas}
\end{center}
\vspace*{-0.75cm}
\end{figure}

%%%%%%%%%%%%%%%%%%%%%%%%%%%%%%%%%%%%%%%%%%%%
As the sum rules reproduce quite well the absolute masses of the (pseudo)scalar states, we can confidently use their mass-spliitngs for extracting $\alpha_s$ and $\la \alpha_s G^2\ra$. We shall not work with the Double Ratio of LSR\,\cite{DRSR,SNFORM1,SNB1,SNB2}

%,SNGh3,
%SNGh1,SNGh5,SNmassa,SNmassb,SNhl,
%HBARYON1,HBARYON2,NAVARRA, SNB1,SNB2} 
as each sum rule does not optimize at the same points. We check that, in the mass-difference , the effect of the choice of the continuum threshold is reduced and induces an error from 6 to 14 MeV instead of 11 to 35 MeV in the absolute value of the masses. The effect due to $\overline{m}_{c,b}$ in Eqs.\,\ref{eq:mcaverage} and \ref{eq:mbaverage} and to $\mu$ in Eqs.\,\ref{eq:muc} and \ref{eq:mub} induce respectively an error of about (1--2) MeV and 8 MeV. The largest effects are due to the changes of $\alpha_s$ and $\la \alpha_s G^2\ra$. We show their correlations in Fig\,\ref{fig:alfas-g2} where we have runned the value of $\alpha_s$ from $\mu=2.85$ GeV to $M_\tau$ in the charm channel and from $\mu=9.5$ GeV to $M_\tau$ in the bottom one where the values of $\mu$ correspond to the scales at which the sum rules have been evaluated:
\bea
&&\hspace*{-1cm} \alpha_s(2.85)=0.262(9) \leadsto\alpha_s(M_\tau)=0.318(15)~\nnb\\
&&\hspace*{0.25cm} \leadsto\alpha_s(M_Z)=0.1183(19)(3) ~,\nnb\\
&&\hspace*{-1cm} \alpha_s(9.50)=0.180(8) \leadsto\alpha_s(M_\tau)=0.312(27)\nnb\\
&&\hspace*{0.25cm} \leadsto\alpha_s(M_Z)=0.1175(32)(3) ~,
\label{eq:alfas1}
%\label{eq:glue1}
\eea
where the last error is due to the running procedure. 
We have requested that the method reproduces within the errors the experimental mass-splittings by about 2-3 MeV. These values are compared  in Fig.\ref{fig:alfas} with the running at different $\mu$ of the world average\,\cite{BETHKEa,PDG}:
\beq
\alpha_s(M_Z)=0.1181(11)~.
\label{eq:asworld}
\eeq
With the central values given in Eqs.\,\ref{eq:g2average} and \,\ref{eq:alfas-mb}, the allowed region  from both charmonium and bottomium channels leads to our final  predictions:
\bea
&&\hspace*{-1cm} \alpha_s(M_\tau)=0.318(15) \leadsto\alpha_s(M_Z)=0.1183(19)(3)~,\nnb\\
&&\hspace*{-1cm}\la \alpha_s G^2\ra=(6.34\pm 0.39)\times 10^{-2}~{\rm GeV}^4~,
\label{eq:alfas}
\label{eq:glue1}
\eea
Adding into the analysis the range of input $\alpha_s$ values given in Eq.\,\ref{eq:param} (light grey horizontal band in Fig.\,\ref{fig:alfas-g2}), one can deduce 
stronger constraints on the value of $\la\alpha_s G^2\ra$:
\beq
\la \alpha_s G^2\ra=(6.39\pm 0.35)\times 10^{-2}~{\rm GeV}^4.
\label{eq:glue2}
\eeq
Combining the previous values in Eqs.\,\ref{eq:respsi}, \ref{eq:glue1} and \ref{eq:glue2} with the ones  in Table\,\ref{tab:g2}, one obtains the {\it new sum rule average}:
\beq
\la \alpha_s G^2\ra\vert_{\rm average}=(6.35\pm 0.35)\times 10^{-2}~{\rm GeV}^4,
\label{eq:glue3}
\eeq
where we have retained the error from the most precise determination in Eq.\,\ref{eq:glue2} instead of the weighted error of 0.23. This result 
definitely rules out some eventual lower and negative values quoted in Table\,\ref{tab:g2}. 
 %%%%%%%%%%%%%%%%%%%%%%%%%%%%%%%%%%%%%%%
%%%%%%%%%%%%%%%%%%%%%%%%%%%%%%%%%%%%%%%%% 
 %%%%%%%%%%%%%%%%%%%%%%%%%%%%%%%%%%%%%%%
%%%%%%%%%%%%%%%%%%%%%%%%%%%%%%%%%%%%%%%%%%
 \section{Correlated values of $\overline{m}_{c,b}(\overline{m}_{c,b})$ from $M_{B_c}$}
 %%%%%%%%%%%%%%%%%%%%%%%%%%%%%%%%%%%%%%%%%%
 We extend the previous analysis to the case of the $B_c$-meson\,\cite{SN19}. We determine simultaneously $\overline{m}_{c}(\overline{m}_{c})$ and $\overline{m}_{b}(\overline{m}_{b})$ from the ratio of sum rules requested to reproduce the $B_c$-mass.
 
\begin{figure}[hbt]
%\vspace*{-0.5cm}
\begin{center}
\includegraphics[width=8.cm]{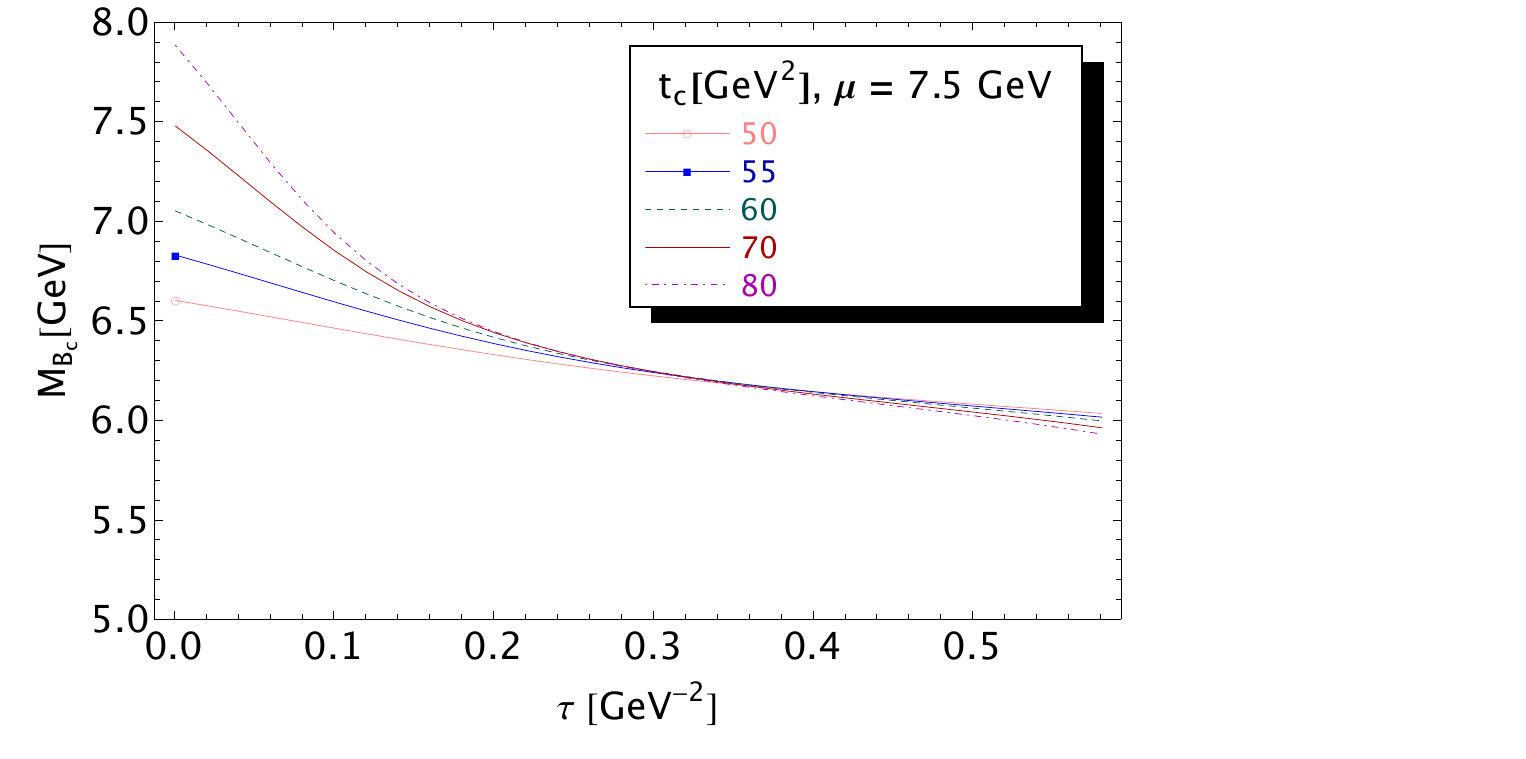}
\vspace*{-0.25cm}
\caption{\footnotesize  $M_{B_c}$ as function of $\tau$ for different values of $t_c$ and for $\mu$=7.5 GeV.} 
\label{fig:mbc-tau}
\end{center}
\vspace*{-0.75cm}
\end{figure}

\b First, we study the $\tau$ and $t_c$-stability of the analysis for given values of $\mu$ and $\overline{m}_{c,b}(\overline{m}_{c,b})$ determined previously. The result is shown in Fig.\,\ref{fig:mbc-tau} where $\tau$ and $t_c$ stabilities are reached for $\tau\simeq 0.3$ GeV$^{-2}$ and $t_c\simeq 50-70$ GeV$^2$. 
 %%%%%%%%%%%%%%%%%%%%%%%%%%%%%%%%%%%%%%%%%
%%%%%%%%%%%%%%%%%%%%%%%%%%%%%%%%%%%%%%%%%

\begin{figure}[hbt]
%\vspace*{-0.5cm}
\begin{center}
\includegraphics[width=8.cm]{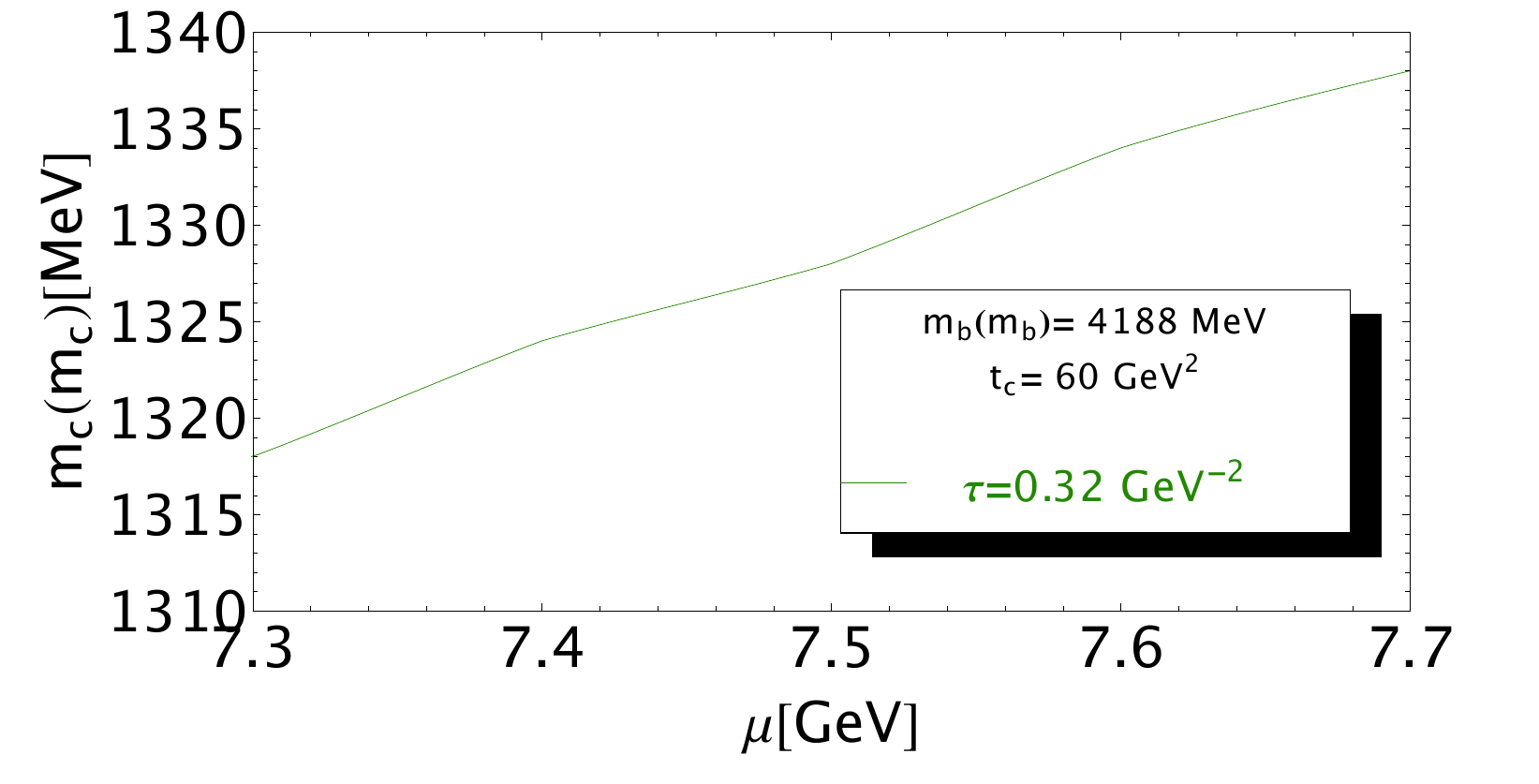}
\vspace*{-0.25cm}
\caption{\footnotesize  ($\overline{m}_c(\overline{m}_c$) values used for reproducing $M_{B_c}$  versus $\mu$ given $\overline{m}_b(\overline{m}_b)=4188$ MeV.} 
\label{fig:mbc-mu}
\end{center}
%\vspace*{-0.75cm}
\end{figure}

\b Second, we study the $\mu$-dependence of the result on Fig.\,\ref{fig:mbc-mu} by fixing $\overline{m}_{b}(\overline{m}_{b})$ and varying the output values of $\overline{m}_{c}(\overline{m}_{c})$ versus $\mu$ where we find a stability for $\mu\simeq (7.5\pm 0.1)$ GeV. 
 %%%%%%%%%%%%%%%%%%%%%%%%%%%%%%%%%%%%%%%
%%%%%%%%%%%%%%%%%%%%%%%%%%%%%%%%%%%%%%%%%%

\begin{figure}[hbt]
\vspace*{-0.25cm}
\begin{center}
\includegraphics[width=8cm]{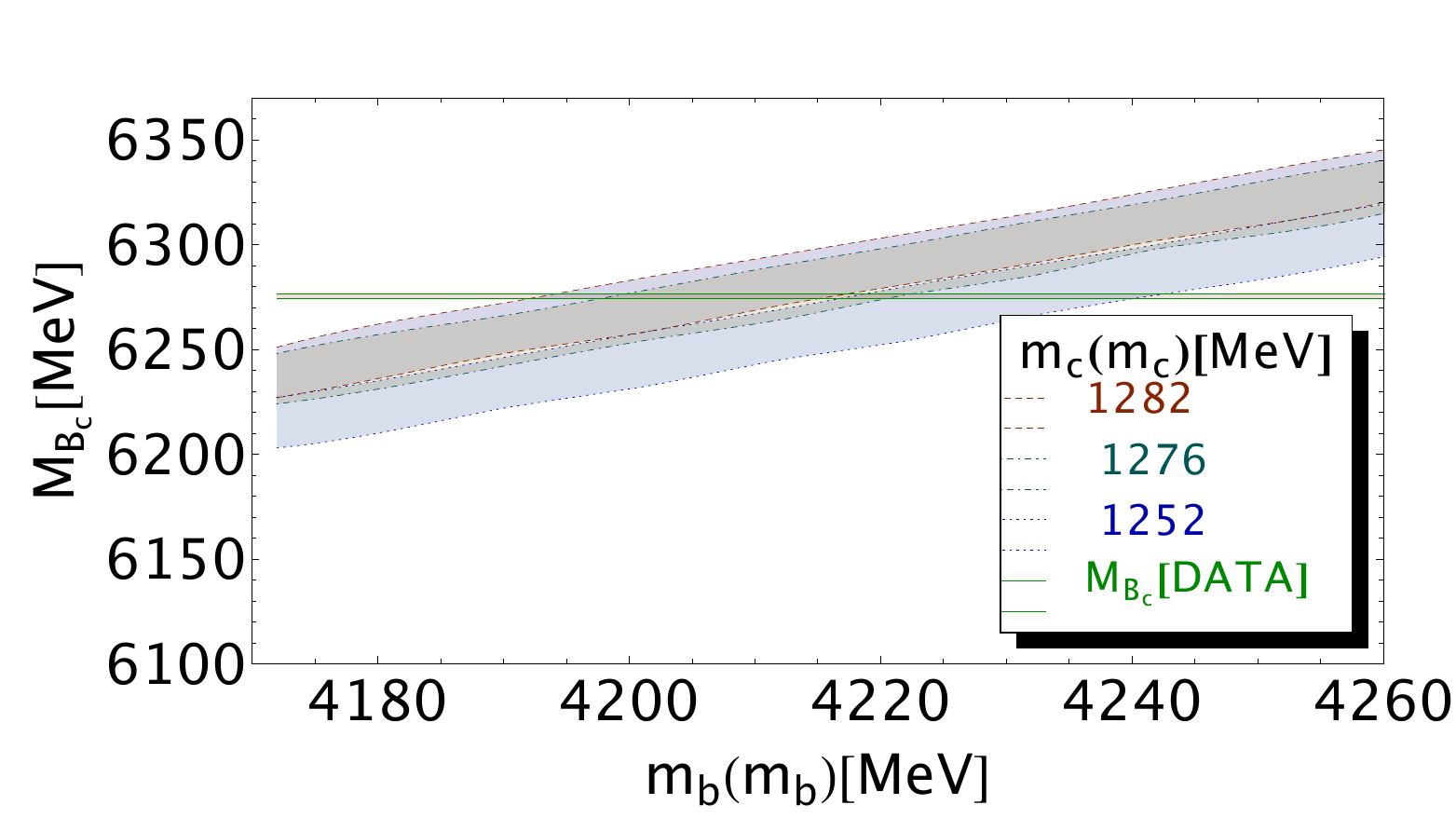}
%\vspace*{-0.25cm}
\caption{\footnotesize  $M_{B_c}$ as function of $\overline{m}_b(\overline{m}_b)$ for different values of  $\overline{m}_c(\overline{m}_c)$ and for $\mu$=7.5 GeV. The band corresponds to the error induced by the localisation of $\tau\simeq (0.30-0.32)$ GeV$^{-2}$. The range of values of $\overline{m}_c(\overline{m}_c)$ and $\overline{m}_b(\overline{m}_b)$ determined from charmonium and bottomium systems using a similar approach are taken from the ones allowed previously, where 
we have multiplied by a factor 2 the quoted error of $\overline{m}_b(\overline{m}_b)$ obtained previously. } 
\label{fig:mbc-mb}
\end{center}
\vspace*{-0.75cm}
\end{figure}

\b Third, given the previous optimal values of $\tau$ and $\mu$, we study the correlation between $\overline{m}_{c}(\overline{m}_{c})$ and $\overline{m}_{b}(\overline{m}_{b})$
by demanding that the sum rule reproduces the $B_c$-mass. 
 The result of the analysis is shown in Fig.\,\ref{fig:mbc-mb}.  
 %%%%%%%%%%%%%%%%%%%%%%%%%%%%%%%%%%%%%%%
We deduce from $M_{B_c}$ and fromthe intersection region allowed by the charmoniumand bottomium sum rules :
\bea
\overline{m}_c(\overline{m}_c)&=&1286(16)\, {\rm MeV} \nnb\\
 \overline{m}_b(\overline{m}_b)&=&4202(7)\,{\rm MeV}. 
\eea
Combined with previous estimates from heavy quarkonia, we deduce the new QCD Spectral Sum Rules (QSSR) tentative average :
\bea
\overline{m}_c(\overline{m}_c)&=&1266(6)\vert_{\rm average}\, {\rm MeV} \nnb\\ \overline{m}_b(\overline{m}_b)&=& 4196(8)\vert_{\rm average}\,{\rm MeV}.
\label{eq:maverage}
\eea
%%%%%%%%%%%%%%%%%%%%%%%%%%%%%%%%%%%%%%%%%%
\section{Decay constants $f_{B_c}$ and $f_{B_c(2S)}$}
\begin{figure}[hbt]
\vspace*{-0.25cm}
\begin{center}
\includegraphics[width=10cm]{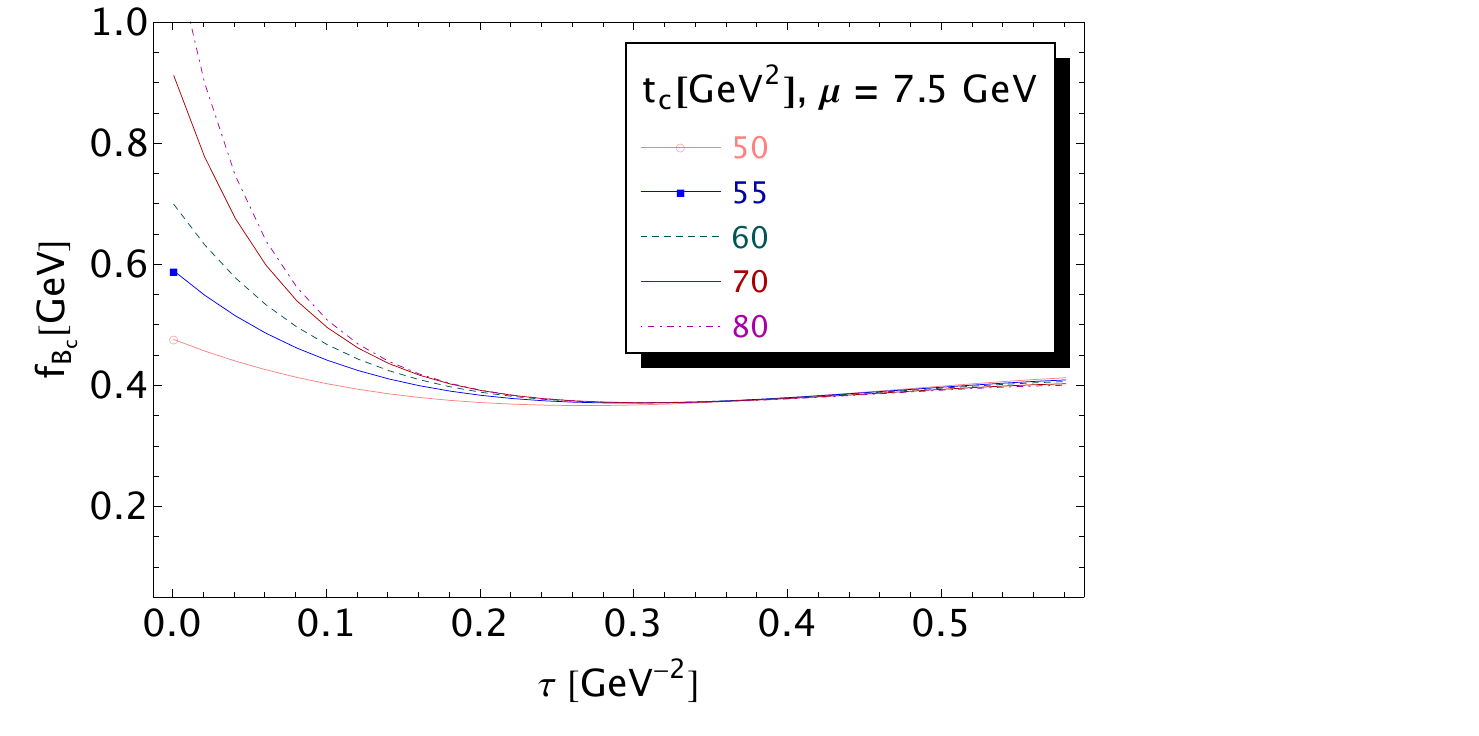}
%\vspace*{-0.25cm}
\caption{\footnotesize  $f_{B_c}$ as function of $\tau$ for different values of $t_c$, for $\mu$=7.5 GeV and for $[\overline{m}_c(\overline{m}_c),\overline{m}_b(\overline{m}_b)]=[1264,4188]$ MeV. } 
\label{fig:fbc-tau}
\end{center}
\vspace*{-0.25cm}
\end{figure}

\begin{figure}[hbt]
\vspace*{-0.25cm}
\begin{center}
\includegraphics[width=8cm]{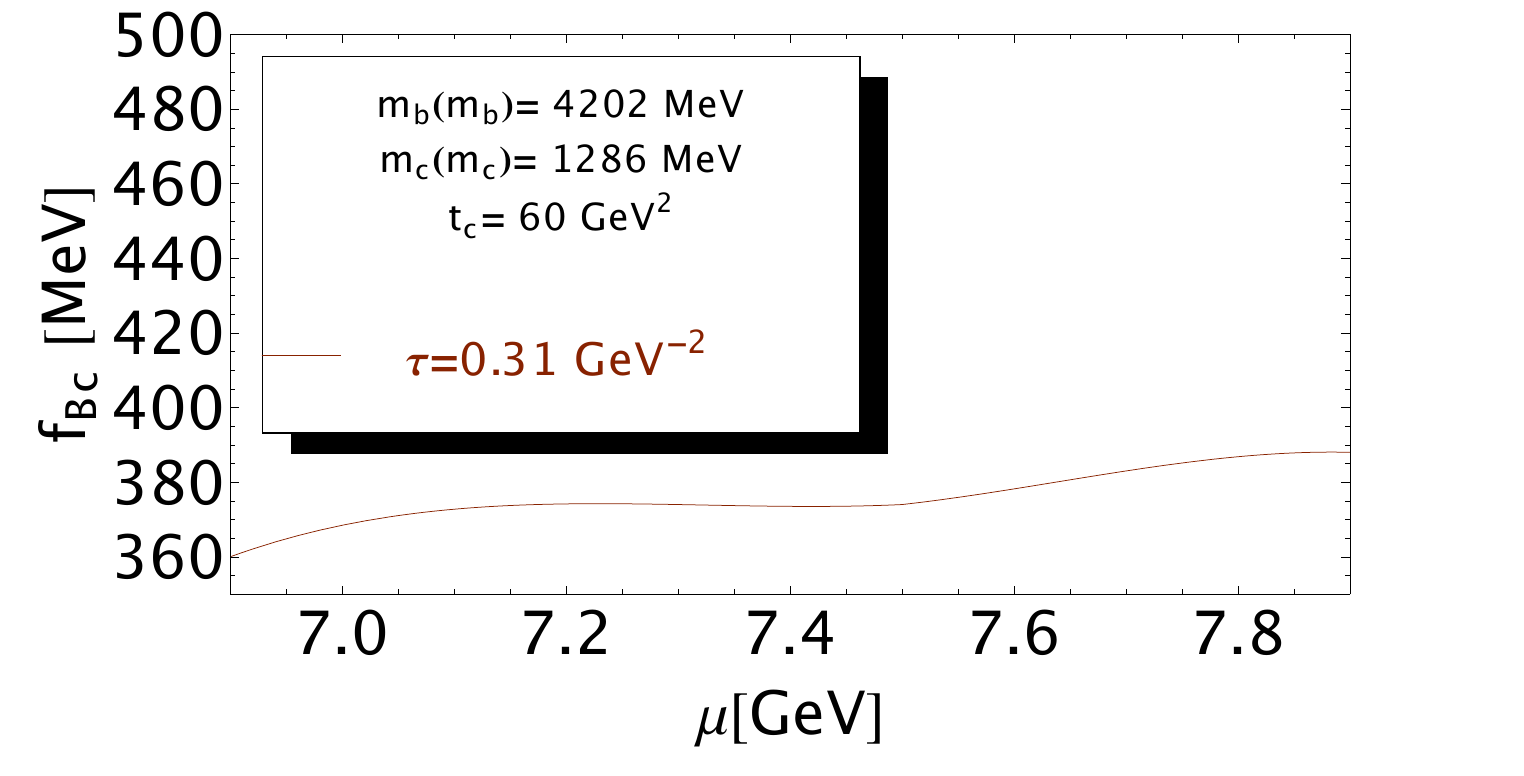}
%\vspace*{-0.25cm}
\caption{\footnotesize  $f_{B_c}$ as function of $\mu$ for $\tau\simeq 0.31$ GeV$^{-2}$ and for and for $[\overline{m}_c(\overline{m}_c),\overline{m}_b(\overline{m}_b)]=[1264,4188]$ MeV.} 
\label{fig:fbc-mu}
\end{center}
\vspace*{-0.75cm}
\end{figure}

%%%%%%%%%%%%%%%%%%%%%%%%%%%%%%%%%%%%%%%%%%
Using the previous correlated values of $\overline{m}_{c,b}(\overline{m}_{c,b})$, we use the LSR ${\cal L}_0^c$ in Eq. 4 for extracting the decay constant $f_{B_c}$ of the $B_c$ meson. The searches for the $(\tau,t_c)$ and $\mu$-stabilities are respectively shown in Figs.\ref{fig:fbc-tau} and \ref{fig:fbc-mu}.  We obtain:
\beq
f_{B_c}=371(17)~{\rm MeV},
\label{eq:fbc}
\eeq
where the largest errors come from the higher order PT corrections. This result confirms previous QSSR results disagrees with the lattice one quoted in Table 3 of Ref.\,\cite{SN14}. 

Using the positivity of the QCD continuum contribution in the spectral function, an upper bound on the $B_c(2S)$ decay constant has been also derived in Ref.\,\cite{SN19}:
\beq
f_{B_c(2S)}\leq 139(6)~{\rm MeV},
\label{eq:fbc2s}
\eeq
where the recent experimental mass $M_{B_c(2S)}=6872(1.5)$ MeV from CMS\,\cite{CMS} has been used. 
 %%%%%%%%%%%%%%%%%%%%%%%%%%%%%%%%%%%%%%%
 %%%%%%%%%%%%%%%%%%%%%%%%%%%%%%%%%%%%%%%
 %%%%%%%%%%%%%%%%%%%%%%%%%%%%%%%%%%%%%%%%%
\section{Summary and Conclusions}
%%%%%%%%%%%%%%%%%%%%%%%%%%%%%%%%%%%%%%%%%

%\hspace*{0.5cm}
\b We have explicitly studied (for the first time) the correlations between $\alpha_s,~\la \alpha_s G^2\ra$ and $\overline{m}_{c,b}$ using ratios of Laplace sum rules @N3LO of PT QCD and including the gluon condensate  $\la \alpha_s G^2\ra$ of dimension 4 @NLO  and the ones of dimension 6-8 @LO  in the (axial-)vector charmonium and bottomium channels.
We have used the criterion of $\mu$-stability in addition to the usual sum rules stability ones (sum rule variable $\tau$ and continuum threshold $t_c$) for extracting our optimal results.  They are given in Eqs.\,\ref{eq:respsi} to \ref{eq:mcaverage} and in Eqs.\,\ref{eq:resmb} and \ref{eq:mbaverage}.

\b We have extended the analysis to the (pseudo)scalar channels where the experimental  masses of the lowest ground states are reproduced quite well. The $\eta_c$ sum rule also leads to an alternative prediction of $\overline{m}_{c}$ in Eq.\,\ref{eq:mcetac}.  

\b Updated average values of the charm and bottom running quark masses from relativistic  QCD spectral sum rules (QSSR) including the new results from $M_{B_c}$ can be respectively found in Eq.\,\ref{eq:maverage}.  

\b These values have been used to extract the value of $f_{B_c}$ quoted in Eq.\ref{eq:fbc} which confirms some previous results quoted in Table 3 of Ref.\,\cite{SN14} but disagrees  with some of them namely the one from Lattice calculations. Upper bound for the $B_c(2S)$ decay constant has been also derived in Eq.\,\ref{eq:fbc2s}.

%The value of $\overline{m}_{b}(\overline{m}_{b})$ agrees within the errors with the less accurate ones from some recent non-relativistic approaches\,\cite{PINEDA,RAUH}.

%\b They are given in Eqs.\,\ref{eq:respsi} to \ref{eq:mcaverage} and in Eqs.\,\ref{eq:resmb} and \ref{eq:mbaverage}.  
%Our results for  $\overline{m}_{c,b}$ confirm our previous estimates from moments in Eq\,\ref{eq:param} and Laplace sum rules in Eq.\,\ref{eq:mblsr} . The results are used for predicting  the (pseudo)scalar lowest ground state masses where there is good agreement with the data  within the errors.  

\b The $\chi_{c0(b0)}-\eta_{c(b)}$ mass-splittings lead to improved values of the gluon condensate $\la \alpha_s G^2\ra$ in Eqs.\,\ref{eq:glue1} and \ref{eq:glue2}.  The new {\it sum rule average}  is given in Eq.\,\ref{eq:glue3} and Table\,\ref{tab:g2}. 

\b Such mass-splittings also provide new predictions of $\alpha_s(\mu)$ at two different scales quoted in Eqs.\,\ref{eq:alfas1} and \ref{eq:alfas} from Fig.\,\ref{fig:alfas-g2}, which are in good agreement with the running of the world average quoted in Eq.\,\ref{eq:asworld} shown in Fig.\,\ref{fig:alfas}\,\footnote{See the discussion in the addendum of Ref.\,\cite{SN18}.}. The most precise prediction given in Eq.\,\ref{eq:alfas}, which we consider as a final estimate from QSSR, comes from the (pseudo)scalar charmonium mass-splittings. 

%%%%%%%%%%%%%%%%%%%%%%%%%%%%%%%%%%%%%%%%%
%\section*{References}
%%%%%%%%%%%%%%%%%%%%%%%%%%%%%%%%%%%%%%%%%
%\newpage
%%%%%%%%%%%
\input{bib_exp19.tex}

%%%%%%%%%%%
\end{document}

%% file: bib_exp19.tex
%%%%%%%%%%%%%%%